# Nonlinear Model Predictive Control-Based Reverse Path-Planning and Path-Tracking Control of a Vehicle with Trailer System

Xincheng Cao, Haochong Chen, Bilin Aksun-Guvenc, Levent Guvenc

Automated Driving Lab, Ohio State University

Brian Link, Peter J Richmond, Dokyung Yim, Shihong Fan, John Harber

HATCI

## Abstract

Reverse parking maneuvers of a vehicle with trailer system is a challenging task to complete for human drivers due to the unstable nature of the system and unintuitive controls required to orientate the trailer properly. This paper hence proposes an optimization-based automation routine to handle the path-planning and path-tracking control process of such type of maneuvers. The proposed approach utilizes nonlinear model predictive control (NMPC) to robustly guide the vehicle-trailer system into the desired parking space, and an optional forward repositioning maneuver can be added as an additional stage of the parking process to obtain better system configurations, before backward motion can be attempted again to get a good final pose. The novelty of the proposed approach is the simplicity of its formulation, as the path-planning and path-tracking operations are only conducted on the trailer being viewed as a standalone vehicle, before the control inputs are propagated to the tractor vehicle via inverse kinematic relationships also derived in this paper. Simulation case studies and hardware-in-the-loop tests are performed, and the results demonstrate the efficacy of the proposed approach.

# 1. Introduction

The development of connected and autonomous or automated vehicles has seen much progress in recent years [1-9] . One of the most important functions of such vehicles is to plan and track their own paths [10], [11]. An extension of this function is the capability of a vehicle to complete path-planning and path-tracking operations with an attached trailer. Reverse parking is one of the most challenging maneuvers to complete for an autonomous or automated vehicle and the difficulty increases considerably in reverse parking of a vehicle with a trailer. The difficulty in this type of maneuver is caused by several reasons. Even for reverse parking of a vehicle alone, an understeer vehicle becomes oversteer when driven in reverse which is usually not a very serious problem as speeds in reverse maneuvering are relatively smaller. This is further complicated for a vehicle with a trailer as the vehicle needs to be steered in the opposite direction of the intended trailer heading which is intuitively difficult for inexperienced drivers, resulting in the use of advisory systems to aid drivers [12]. These differences may result in not being able to follow the desired path for the trailer and oscillatory behavior in reverse motion. Another challenge is the fact that to orientate the trailer the same way, different steering inputs at the vehicle will be required depending on the current system configuration. This motivates the research reported in this paper on automating the vehicle-trailer reverse parking process.

A suitable system model is needed first for the vehicle-trailer combination to automate its reverse parking maneuver. In general, there are two types of models typically used for such systems: dynamic models that consider the forces involved and kinematic models that focus on geometric relationships. The dynamic models can be derived using various approaches. In [13], for example, a tractor-trailer system with a double trailer configuration is analyzed, where one of the trailers is a semi-trailer and the other a full trailer. This setup results in the whole system being treated as a four-component multi-body system. The Newton-Euler approach is subsequently used to derive the system model, which makes use of a linear tire model. Similarly, [14] illustrates the derivation of a dynamic semi-tractor-trailer model using the Newton-Euler method while exploring two different types of tire side slip angle models. The Newton-Euler method is also utilized in [15] to derive the dynamics of a hydraulically-steered articulated vehicle. A similar hydraulically articulated steer vehicle is also studied in [16] and a 12-DOF (degree-of-freedom) dynamic model is derived as a result of that effort. Another method typically used in the derivation of dynamic vehicle-trailer models is the use of Euler-Lagrange equations. For instance, 17] provides the derivation procedure for the dynamic models of articulated commercial vehicles with various configurations using the Euler-Lagrange method. Similarly, [18] derives the dynamic model of an articulated truck using the same approach, where the system is treated as a five-axle articulated bicycle. Furthermore, [19] arrives at a nonlinear 4-DOF model using the Lagrange equations and subsequently shrinks the DOF from 4 to 3 through model reduction techniques. Some comparative studies also exist to evaluate the accuracy of dynamic models. For instance, [20] compares the models described in [13], [14], [17], [18] under the five-axle configuration, while [21] compares four different dynamic vehicle-trailer models: linear 3-DOF, nonlinear 4-DOF, nonlinear 6-DOF and a high-fidelity nonlinear 21-DOF CarSim model. Aside from dynamic models, plenty of literature also covers the derivation of simpler kinematic models to analyze the behaviors of a vehicle-trailer system, particularly at low speed where parking maneuvers are typically carried out, as significant tire deformations do not typically occur. Reference [22] describes the procedure to derive a kinematic vehicle-trailer model for a one-trailer configuration, partially making use of the instantaneous center of rotation for the vehicle unit. Reference [23], on the other hand, uses kinematic constraint equations to derive the kinematic model for a system containing a tractor, a semi-trailer and a drawbar trailer. The same method, based on kinematic constraint equations, is also utilized in [24] to arrive at the kinematic model of a system consisting of a semi-tractor and a semi-trailer. Meanwhile, [25] demonstrates the kinematic model of a generic n-trailer setup while discussing the flatness properties of such a system.

Due to the multi-body nature of vehicle-trailer systems, effective control of such systems can prove to be challenging, especially if the system is to travel in reverse, where its behavior may become unstable. As a result, one common strategy in the existing literature is to use two-stage controllers. For instance, [26] and [27] present a high-level controller that generates desired hitch angle and a low-level controller to follow such an angle. Similarly, [28] describes a high-level controller that generates desired curvature and a low-level controller to track the curvature requirement. Additionally, [29] demonstrates a control system that first calculates the expected tractor and trailer yaw rates before generating a steering input to follow them. Another type of two-stage approach regards the last trailer unit in the vehicle-trailer system chain as a 'virtual tractor', and the required steer angle at this last trailer for its proper orientation is subsequently translated to the steerable axle of the tractor through geometric relations. References [12], [23], [30], [31] describe several variants of this method in detail. A further variation of this 'virtual tractor' approach is to introduce an additional 'virtual tractor' behind the last trailer unit in the system chain, where its steering axle is located at the path-tracking preview point when the system is travelling in reverse. Reference [24] discusses this approach in detail. Apart from the above-mentioned two-stage control schemes, other approaches also exist in the literature. For example, [32] decouples a kinematic tractor-trailer system into two subsystems as the tractor component and the trailer component and uses distributed continuous time optimal control to achieve path-tracking functionality in the forward direction of motion, Reference [33], on the other hand, uses a nonlinear model-reference control system for tractor-trailer system stabilization with results for the forward direction of motion. The solution presented in this paper uses a one stage nonlinear model predictive controller that can handle path planning and path tracking simultaneously thereby reducing the complexity of the two stage control approach.

In order to perform meaningful path-tracking control, a collision-free feasible path should first be generated. The challenge of such a path-planning procedure is two-fold: the path must be kinetically/dynamically feasible for the vehicle-trailer system in question, and the path should be collision-free for both the vehicle and trailer units. Due to the possibly complicated nature of the collision-free feasible path, some approaches opt to plan the path that starts from the terminal/goal position and ends at the starting point. For example, [34] introduces a 'cascade path-planning' routine for the collision-free path planning of a tractor-trailer system represented by a kinematic motion model. Similarly, [35] also tackles the docking problem of the tractor-trailer system starting from the terminal state, where an initial path is first generated from parked position to a point close to the starting position using a tree-based path planner, and an optimization problem is then applied to mimic the initial path so that a collision-free, kinematically feasible optimal path that takes the system from the initial position to the terminal position can be generated. It should be remarked that both [34] and [35] use an of optimization-based method at some stage of the procedure. In fact, optimization-based approaches are very popular in the literature for feasible path-planning purposes. For example, [36] presents a Pontryagin's Minimum Principle (PMP) based optimal path-planning routine for the tractor-trailer system that can perform obstacle avoidance through artificial potential field in the forward direction of motion. Reference [37], on the other hand, presents a lattice-based path planner that utilizes optimization to generate kinematically feasible motion primitives. Apart from optimization-based approaches, some other path-planning techniques and results exist as well. For example, [38] provides two path-planners for the tractor-trailer system: an exact local path planner for the simple one-trailer system by exploiting the differential flatness property of such a system, and an RRT-based motion planner. Meanwhile, [39] discusses the minimum parking space required for the tractor-trailer system with one-axle trailer to attempt the parallel parking maneuver. Reference [40], on the other hand, presents an exact motion planner that constructs feasible paths for the tractor-trailer system by concatenating simple path constructs such as rotations, translations, stretches and bends. Reference [41] uses semi-supervised learning, where a deep neural network is used to generate paths that minimize off-track of the area swept by the tractor-trailer system. Additionally, [42] proposes a cooperative trajectory planning algorithm for tractor-trailer wheeled robots, where motion characteristics of both the tractor and the trailer are considered.

The above-mentioned path-planning approaches, while effective, are not sufficiently simple in formulation. This paper hence proposes a nonlinear model predictive control (NMPC) path-planning and control scheme that is simple in construction while also providing closed-loop control actions that are segmented in nature and can be applied in a finite number of steps. This is significant as the reverse parking of a vehicle and a vehicle with a trailer is a short-distance problem where a driver typically executes steering changes in a discrete fashion with large changes between steps. The MPC method is ideal to generate such control action and fits the short distance parking problem very naturally. The outline of the rest of this paper is as follows. Section 2 details the derivation of a kinematic vehicle-trailer model. Section 3 illustrates the derivation of inverse kinematics of the vehicle-trailer system. Section 4 develops the nonlinear model predictive control (NMPC) based reverse parking path-planning design. Section 5 subsequently explains the design of the forward repositioning maneuver. Finally, Section 6 presents simulation case study results. The paper ends with conclusions.

## 2. Kinematic Vehicle-Trailer Model

In order to automate the vehicle-trailer backup parking maneuver, a system model must first be derived. Given the low-speed nature of the vehicle-trailer system during reverse motion, a kinematic model is ideal for this purpose as it combines reasonable accuracy with simple formulation. This section aims to derive such a kinematic model.

The schematic of the generic vehicle-trailer system for the one-trailer case is displayed in Figure 1, and the parameters of this model are listed in Table 1. Note that value selection of $L_H$ which is the distance from the vehicle rear axle to the hitch joint determines the type of tractor vehicle (semi-tractor or passenger vehicle), and its value can be negative (although not desirable) in the semi-tractor case. In the case of this paper, however, the value of $L_H$ is assumed to be positive, as the aim is to explore the automated reverse parking maneuvers of a vehicle-trailer system with a car, SUV or pickup truck type tractor unit with a hitch connected to the rear of the vehicle. A front wheel steering vehicle is considered in this paper.

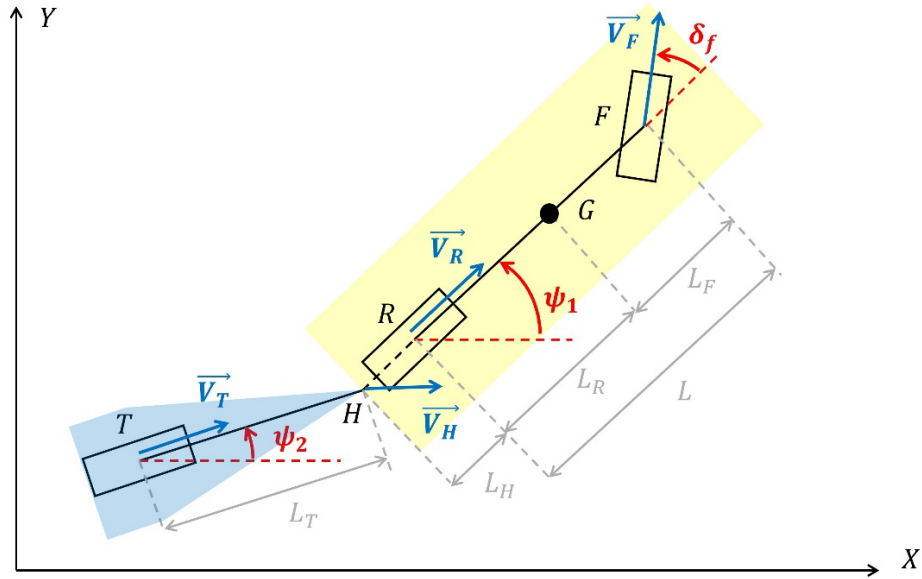

**Figure 1.** Kinematic vehicle-trailer model with one trailer

**Table 1.** Parameters of kinematic tractor-trailer model

| Model Parameter | Explanation |
|---|---|
| $L$ | Wheelbase of the tractor vehicle (car, SUV or pickup truck) |
| $L_F$ | Distance between vehicle center of gravity G and front axle |
| $L_R$ | Distance between vehicle center of gravity G and rear axle |
| $L_H$ | Distance between vehicle rear axle and trailer hitch joint |
| $L_T$ | Distance between trailer axle and trailer hitch joint |
| $\delta_f$ | Vehicle front wheel steer angle |
| $\psi_1$ | Vehicle yaw angle |
| $\psi_2$ | Trailer yaw angle |
| $\vec{V_F}$ | Vehicle front axle velocity |
| $\vec{V_R}$ | Vehicle rear axle velocity |
| $\vec{V_H}$ | Trailer hitch velocity |
| $\vec{V_T}$ | Trailer axle velocity |

If we assign the vehicle rear axle as the position reference point, we can model the vehicle and trailer as rigid bodies moving in the X-Y plane that are connected at the hinge joint and derive the position equations of key points as shown below in Equations (1) to (6).

$$X_F = X_R + L \cdot \cos(\psi_1) \qquad (1)$$

$$Y_F = Y_R + L \cdot \sin(\psi_1) \tag{2}$$

$$X_H = X_R - L_H \cdot \cos(\psi_1) \tag{3}$$

$$Y_H = Y_R - L_H \cdot \sin(\psi_1) \tag{4}$$

$$X_T = X_H - L_T \cdot \cos(\psi_2) \tag{5}$$

$$Y_T = Y_H - L_T \cdot \sin(\psi_2) \tag{6}$$

$X_R$ and $Y_R$ are the coordinates of the rear axle center. $X_F$ and $Y_F$ are the coordinates of the front axle center. $X_H$ and $Y_H$ are the coordinates of the hinge joint. $X_T$ and $Y_T$ are the coordinates of the trailer axle center. Further applying time derivatives to the position equations above yields velocity equations as displayed in Equations (7) to (10).

$$\vec{V_R} = \begin{pmatrix} \dot{X}_R \\ \dot{Y}_R \end{pmatrix} \tag{7}$$

$$\vec{V_F} = \begin{pmatrix} \dot{X}_F \\ \dot{Y}_F \end{pmatrix} = \begin{pmatrix} \dot{X}_R - L \cdot \dot{\psi}_1 \cdot \sin(\psi_1) \\ \dot{Y}_R + L \cdot \dot{\psi}_1 \cdot \cos(\psi_1) \end{pmatrix} \tag{8}$$

$$\vec{V_H} = \begin{pmatrix} \dot{X}_H \\ \dot{Y}_H \end{pmatrix} = \begin{pmatrix} \dot{X}_R + L_H \cdot \dot{\psi}_1 \cdot \sin(\psi_1) \\ \dot{Y}_R - L_H \cdot \dot{\psi}_1 \cdot \cos(\psi_1) \end{pmatrix} \tag{9}$$

$$\vec{V_T} = \begin{pmatrix} \dot{X}_T \\ \dot{Y}_T \end{pmatrix} = \begin{pmatrix} \dot{X}_H + L_T \cdot \dot{\psi}_2 \cdot \sin(\psi_2) \\ \dot{Y}_H - L_T \cdot \dot{\psi}_2 \cdot \cos(\psi_2) \end{pmatrix} \tag{10}$$

$\vec{V_R}$ is the velocity of the rear axle center. $\vec{V_F}$ is the velocity of the front axle center. $\vec{V_H}$ is the velocity of the hinge joint. $\vec{V_T}$ is the velocity of the trailer axle center.

It can be observed in Figure 1 that the orientations of the tires in the system are aligned with the directions of their velocity vectors. This is due to the assumption that at low speed typical of parking maneuvers, there is minimal tire deformation, meaning that the direction of tire motion follows the direction in which the tire is pointing. This tire no side slip assumption will induce kinematic constraints in the model, and they can be represented in Equations (11) to (13).

$$\tan(\psi_1) = \frac{V_R \cdot \sin(\psi_1)}{V_R \cdot \cos(\psi_1)} = \frac{\dot{Y}_R}{\dot{X}_R} \tag{11}$$

$$\tan(\psi_1 + \delta_f) = \frac{\dot{Y}_F}{\dot{X}_F} \tag{12}$$

$$\tan(\psi_2) = \frac{V_T \cdot \sin(\psi_2)}{V_T \cdot \cos(\psi_2)} = \frac{\dot{Y}_T}{\dot{X}_T} \tag{13}$$

It should be remarked that the term $V_R$ is the magnitude of the vehicle rear axle center velocity $\vec{V_R}$ and can either be positive or negative. Similarly, $V_T$ is the magnitude of the trailer axle center velocity $\vec{V_T}$.

Combining the velocity equations with the kinematic constraint equations and simplifying, one can obtain Equations (14) and (15).

$$\tan(\psi_1 + \delta_f) = \frac{V_R \cdot \sin(\psi_1) + L \cdot \dot{\psi}_1 \cdot \cos(\psi_1)}{V_R \cdot \cos(\psi_1) - L \cdot \dot{\psi}_1 \cdot \sin(\psi_1)} \tag{14}$$

$$\tan(\psi_2) = \frac{V_T \cdot \sin(\psi_2)}{V_T \cdot \cos(\psi_2)} = \frac{V_R \cdot \sin(\psi_1) - L_H \cdot \dot{\psi}_1 \cdot \cos(\psi_1) - L_T \cdot \dot{\psi}_2 \cdot \cos(\psi_2)}{V_R \cdot \cos(\psi_1) + L_H \cdot \dot{\psi}_1 \cdot \sin(\psi_1) + L_T \cdot \dot{\psi}_2 \cdot \sin(\psi_2)} \tag{15}$$

Simplifying individual components in Equations (14) and (15) further, one can obtain Equations (16) to (18), where the yaw behaviors of the vehicle and the trailer as well as the magnitude of trailer axle velocity are expressed.

$$\dot{\psi}_1 = \frac{V_R}{L} \tan(\delta_f) \tag{16}$$

$$\dot{\psi}_2 = \frac{V_R}{L_T}[\sin(\Delta\psi) - \frac{L_H}{L}\cos(\Delta\psi)\tan(\delta_f)] \tag{17}$$

$$V_T = V_R[\cos(\Delta\psi) + \frac{L_H}{L}\sin(\Delta\psi)\tan(\delta_f)] \tag{18}$$

Note that $\Delta\psi = \psi_1 - \psi_2$ is defined as the hitch angle and represents the relative angle between vehicle and trailer orientations. One can then reorganize existing derivation results and formulate the vehicle-trailer kinematic model as displayed in Equations (19) to (24).

$$\dot{X}_R = V_R \cdot \cos(\psi_1) \tag{19}$$

$$\dot{Y}_R = V_R \cdot \sin(\psi_1) \tag{20}$$

$$\dot{\psi}_1 = \frac{V_R}{L} \tan(\delta_f) \tag{21}$$

$$\dot{X}_T = V_R \cos(\psi_2)[\cos(\Delta\psi) + \frac{L_H}{L}\sin(\Delta\psi)\tan(\delta_f)] \tag{22}$$

$$\dot{Y}_T = V_R \sin(\psi_2)[\cos(\Delta\psi) + \frac{L_H}{L}\sin(\Delta\psi)\tan(\delta_f)] \tag{23}$$

$$\dot{\psi}_2 = \frac{V_R}{L_T}[\sin(\Delta\psi) - \frac{L_H}{L}\cos(\Delta\psi)\tan(\delta_f)] \tag{24}$$

In this model, the inputs are the vehicle front axle steer angle $\delta_f$ and vehicle rear axle speed $V_R$.

## 3. Inverse Kinematics

### 3.1. Inverse Kinematics Derivation

One of the main difficulties of vehicle-trailer reverse maneuvers comes from the fact that the system tends to demonstrate some very unintuitive yaw behaviors for inexperienced drivers. As a result, it would be helpful to first regard the trailer as a standalone vehicle and calculate a 'virtual' steering angle at the trailer that would orientate it properly, before mapping this 'virtual' angle to the actual steer angle at the vehicle steerable axle through kinematic derivation. This section illustrates the procedure to derive such a mapping relationship between the 'virtual' steer angle at the trailer and the actual steer angle at the vehicle.

The inverse kinematic vehicle-trailer model under consideration is identical to the model illustrated in Figure 1 except for the addition of a 'virtual' steerable axle at the trailer hitch, as illustrated in Figure 2 that shows only the trailer part of the model to reduce visual clutter. With this virtual steerable axle, the trailer unit can be regarded as a standalone vehicle and its 'virtual' steer angle is denoted as $\delta_T$.

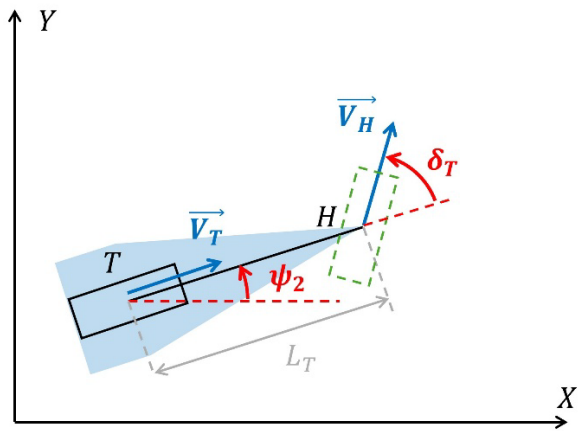

**Figure 2.** Inverse kinematic vehicle-trailer model: trailer part

The procedure to derive this inverse kinematic model is also like that outlined in Section 2, except that we now regard the trailer axle as the position refence point. As a result, one can obtain the position Equations (25) to (30).

$$X_H = X_T + L_T \cdot \cos(\psi_2) \quad (25)$$

$$Y_H = Y_T + L_T \cdot \sin(\psi_2) \quad (26)$$

$$X_R = X_H + L_H \cdot \cos(\psi_1) \quad (27)$$

$$Y_R = Y_H + L_H \cdot \sin(\psi_1) \quad (28)$$

$$X_F = X_R + L \cdot \cos(\psi_1) \quad (29)$$

$$Y_F = Y_R - L \cdot \sin(\psi_1) \quad (30)$$

and the velocity Equations (31) to (34).

$$\vec{V_T} = \begin{pmatrix} \dot{X}_T \\ \dot{Y}_T \end{pmatrix} \quad (31)$$

$$\vec{V_H} = \begin{pmatrix} \dot{X}_H \\ \dot{Y}_H \end{pmatrix} = \begin{pmatrix} \dot{X}_T - L_T \cdot \dot{\psi}_2 \cdot \sin(\psi_2) \\ \dot{Y}_T + L_T \cdot \dot{\psi}_2 \cdot \cos(\psi_2) \end{pmatrix} \quad (32)$$

$$\vec{V_R} = \begin{pmatrix} \dot{X}_R \\ \dot{Y}_R \end{pmatrix} = \begin{pmatrix} \dot{X}_H - L_H \cdot \dot{\psi}_1 \cdot \sin(\psi_1) \\ \dot{Y}_H + L_H \cdot \dot{\psi}_1 \cdot \cos(\psi_1) \end{pmatrix} \quad (33)$$

$$\vec{V_F} = \begin{pmatrix} \dot{X}_F \\ \dot{Y}_F \end{pmatrix} = \begin{pmatrix} \dot{X}_R - L \cdot \dot{\psi}_1 \cdot \sin(\psi_1) \\ \dot{Y}_R + L \cdot \dot{\psi}_1 \cdot \cos(\psi_1) \end{pmatrix} \quad (34)$$

Also similar to Section 2, the kinematic constraint equations can be obtained in Equations (35) to (38) by applying tire no side slip condition.

$$\tan(\psi_2) = \frac{V_T \cdot \sin(\psi_2)}{V_T \cdot \cos(\psi_2)} = \frac{\dot{Y}_T}{\dot{X}_T} \quad (35)$$

$$\tan(\psi_2 + \delta_T) = \frac{\dot{Y}_H}{\dot{X}_H} \quad (36)$$

$$\tan(\psi_1) = \frac{\dot{Y}_R}{\dot{X}_R} \quad (37)$$

$$\tan(\psi_1 + \delta_f) = \frac{\dot{Y}_F}{\dot{X}_F} \quad (38)$$

It should be remarked that the kinematic constraint at the trailer virtual steer axle also applies, meaning that the trailer hitch velocity vector $\vec{V_H}$ is aligned with the 'virtual' steer angle $\delta_T$. Combining velocity equations with kinematic constraint equations, one can obtain Equations (39) to (41).

$$\tan(\psi_2 + \delta_T) = \frac{V_T \cdot \sin(\psi_2) + L_T \cdot \dot{\psi}_2 \cdot \cos(\psi_2)}{V_T \cdot \cos(\psi_2) - L_T \cdot \dot{\psi}_2 \cdot \sin(\psi_2)} \quad (39)$$

$$\tan(\psi_1) = \frac{V_R \cdot \sin(\psi_1)}{V_R \cdot \cos(\psi_1)} = \frac{V_T \cdot \sin(\psi_2) + L_T \cdot \dot{\psi}_2 \cdot \cos(\psi_2) + L_H \cdot \dot{\psi}_1 \cdot \cos(\psi_1)}{V_T \cdot \cos(\psi_2) - L_T \cdot \dot{\psi}_2 \cdot \sin(\psi_2) - L_H \cdot \dot{\psi}_1 \cdot \sin(\psi_1)} \quad (40)$$

$$\tan(\psi_1 + \delta_f) = \frac{V_T \cdot \sin(\psi_2) + L_T \cdot \dot{\psi}_2 \cdot \cos(\psi_2) + L_H \cdot \dot{\psi}_1 \cdot \cos(\psi_1) + L \cdot \dot{\psi}_1 \cdot \cos(\psi_1)}{V_T \cdot \cos(\psi_2) - L_T \cdot \dot{\psi}_2 \cdot \sin(\psi_2) - L_H \cdot \dot{\psi}_1 \cdot \sin(\psi_1) - L \cdot \dot{\psi}_1 \cdot \sin(\psi_1)} \quad (41)$$

Further simplifications of individual components in Equations (39) to (41) yields Equations (42) to (45).

$$\dot{\psi}_2 = \frac{V_T}{L_T}\tan(\delta_T) \tag{42}$$

$$\dot{\psi}_1 = \frac{V_T}{L_H}[\sin(\Delta\psi) - \cos(\Delta\psi)\tan(\delta_T)] \tag{43}$$

$$V_T = \frac{V_R}{\cos(\Delta\psi) + \sin(\Delta\psi)\tan(\delta_T)} \tag{44}$$

$$\delta_f = \mathrm{atan}\left(\frac{L}{L_H} \cdot \frac{\sin(\Delta\psi) - \cos(\Delta\psi)\tan(\delta_T)}{\cos(\Delta\psi) + \sin(\Delta\psi)\tan(\delta_T)}\right) \tag{45}$$

Note that Equations (42) and (43) represent the 'desired' yaw rates of the trailer and the vehicle, respectively, given a virtual steer angle $\delta_T$. Meanwhile, Equation (44) represents the mapping from vehicle rear axle center speed to trailer axle center speed given a virtual steer angle $\delta_T$. Equation (45), on the other hand, is the mapping equation from the virtual steer angle at the trailer hitch to the actual steer angle at the vehicle steerable (front) axle.

### 3.2. Inverse Kinematics Validation

A simulation study is performed to demonstrate the effects of the actual-virtual steering angle mapping equation derived in Section 3.1. The simulation routine is illustrated in Figure 3. A 'desired' $\delta_T$ profile is first generated and fed into the inverse kinematics calculation block that invokes the actual-virtual steering angle mapping equation, and the resulting vehicle steer angle required is plugged into the vehicle-trailer model. An additional block is also included to calculate the 'actual' $\delta_T$ based on the outputs of the vehicle-trailer model, which is then compared to the 'desired' $\delta_T$ profile. The calculation procedure of the 'actual' $\delta_T$ is detailed in Equation (46).

$$Actual\ \delta_T = \mathrm{atan}\left(\frac{\dot{Y}_H}{\dot{X}_H}\right) - \psi_2 = \mathrm{atan}\left(\frac{V_T\sin(\psi_2) + L_T\dot{\psi}_2\cos(\psi_2)}{V_T\cos(\psi_2) - L_T\dot{\psi}_2\sin(\psi_2)}\right) - \psi_2 \tag{46}$$

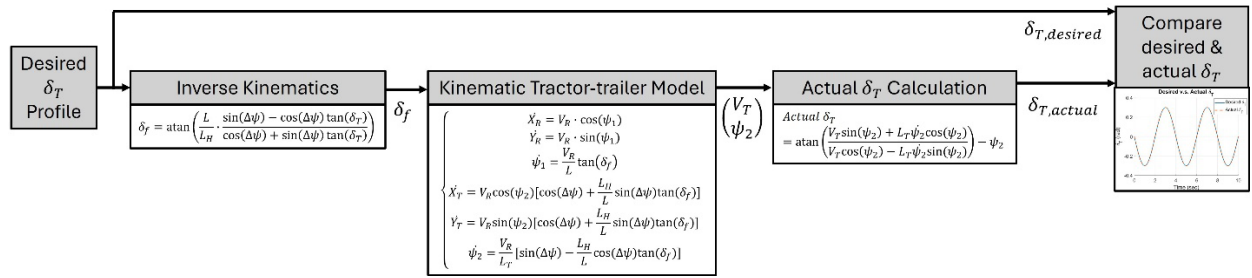

**Figure 3.** Inverse kinematics simulation study model structure

The parameter value choices used in this simulation study are listed in Table 2. It is worth remarking that this simulation study focuses on the case of reverse motion with the trailer hitch located behind the vehicle rear axle, as this paper limits its scope on vehicle-trailer system backup maneuvers with a car-like (car, SUV or pickup truck) tractor unit. Figure 4 displays the result of the simulation. It can be observed that the vehicle steering inputs generated by the inverse kinematics calculation can accurately re-create the desired virtual steering angle profile.

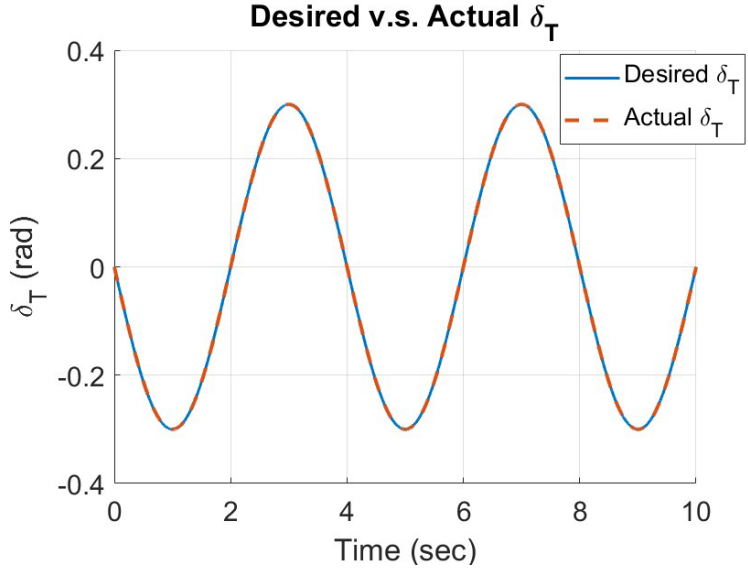

**Figure 4.** Desired virtual steer angle tracking performance for trailer hitch behind vehicle rear axle in reverse motion

**Table 2.** Parameter value choices for inverse kinematics simulation study

| Model Parameter | Value Choice |
| --- | --- |
| $L$ | 3 [m] |
| $L_H$ | 1 [m] (passenger vehicle) |
| $L_T$ | 2.5 [m] |
| $V_R$ | -1 [m/s] (backward motion) |

## 4. Nonlinear Model Predictive Control-Based Path-Planning Design

### 4.1. Nonlinear Model Predictive Control Formulation

In order to properly park a vehicle-trailer combination, one must accomplish both of the following two tasks: 1) generate a feasible path that allows the system to move from its initial state into the parking space; 2) generate a control sequence/law to follow the planned path. Model predictive control (MPC) is a popular approach for this purpose as it can achieve both goals simultaneously. Moreover, while not being able to generate a globally optimal solution due to its finite horizon planning nature, MPC offers closed-loop control laws that are more robust as the optimization is re-run constantly with updated states. The finite horizon nature of MPC is suitable for parking in reverse motion which involves short distance and hence a short and finite time horizon. This section proposes a nonlinear model predictive control (NMPC) formulation to tackle the vehicle-trailer system reverse parking problem, where the vehicle-trailer combination needs to dock into a parking space with designated position and orientation. Figure 5 shows an example of such a problem setup, where it is desirable for the trailer unit to reach zero states when the parking maneuver concludes. It should be remarked that the X-Y coordinate system presented in Figure 5 is identical to that shown in Figure 1 and is initialized to the parking space before the path-planning procedure begins.

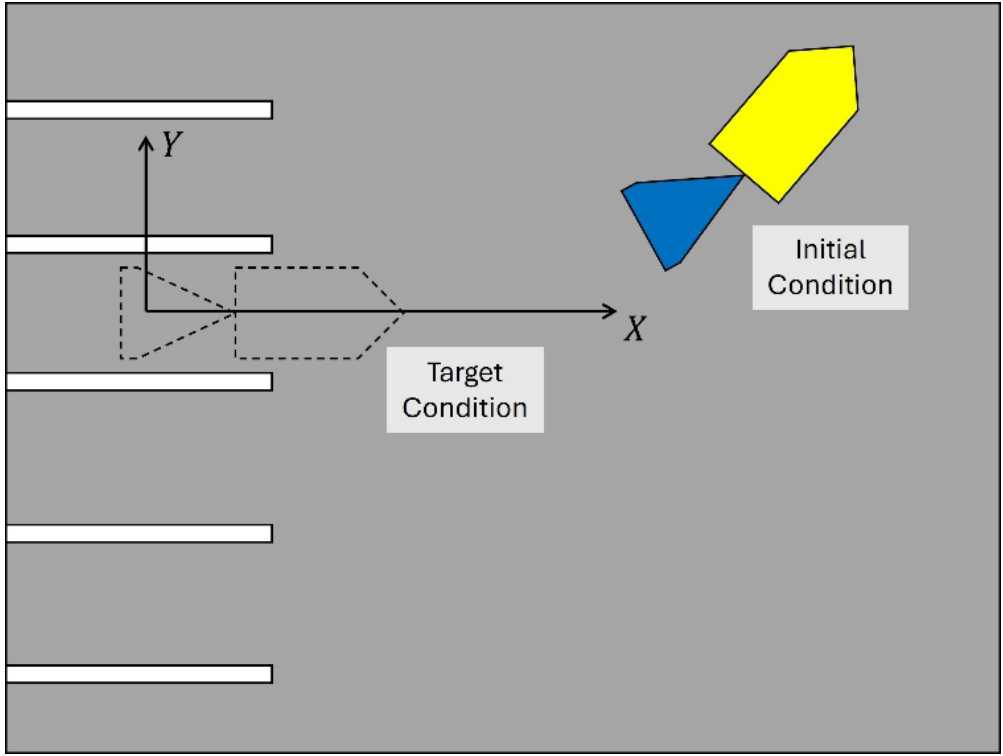

**Figure 5.** An example of vehicle-trailer system reverse parking problem setup

In order to reduce the dimension of the optimization problem, path planning is only performed for the standalone trailer unit with a virtual steering axle at the trailer hitch, as shown in Figure 2. The planning solution is then propagated to the vehicle unit through inverse kinematics derived in Section 3 to generate actual control inputs. As a result, the kinematic model used in this NMPC formulation only contains trailer position and orientation as its states and can be written as shown in Equation (47), where trailer axle speed and virtual steering angle at the trailer hitch serve as the inputs to be optimized.

$$\dot{X} = \begin{bmatrix} \dot{X}_T \\ \dot{Y}_T \\ \dot{\psi}_2 \end{bmatrix} = f(X, U) = \begin{bmatrix} V_T \cdot \cos(\psi_2) \\ V_T \cdot \sin(\psi_2) \\ \frac{V_T}{L_T} \tan(\delta_T) \end{bmatrix}, \text{ where } \begin{cases} X = \begin{bmatrix} X_T \\ Y_T \\ \psi_2 \end{bmatrix} \epsilon \mathbb{R}^{3 \times 1} \\ U = \begin{bmatrix} V_T \\ \delta_T \end{bmatrix} \epsilon \mathbb{R}^{2 \times 1} \end{cases} \quad (47)$$

Since MPC-based approaches only work with discrete models, a discretization routine must be applied to the continuous model described in Equation (47). Equation (48) provides an example of this step using Euler-Forward method with a user-defined time step $d_t$.

$$X_{k+1} = X_k + d_t \cdot f(X, U) = g(X_k, U_k) \qquad (48)$$

One can then define the cost function for the optimization problem. Equation (49) provides a quadratic cost function formulation where the path-planning goal for the trailer unit is to reach zero states (position and orientation) at the terminal time.

$$J = \sum_{k=0}^{N-1}(X_k^T Q X_k + U_k^T R U_k) + X_N^T P X_N, \text{ where } \begin{cases} Q \in \mathbb{R}^{3\times 3} > 0 \\ R \in \mathbb{R}^{2\times 2} > 0 \\ P \in \mathbb{R}^{3\times 3} > 0 \end{cases} \qquad (49)$$

$N$ is the prediction/control horizon and is user-defined as well. As previously shown in Figure 5, the X-Y coordinate system origin is set to the desired final location for the trailer axle center with the X-axis being aligned with the final desired trailer orientation. $X_N$ denotes the final state at the end of the horizon which is desired to be zero. $Q$ and $P$ matrices are positive definite matrices that penalize the trailer states for not being the desired ones. Meanwhile, $R$ is another positive definite matrix that penalizes control inputs with large magnitudes.

With the discrete kinematics and the cost function defined, the optimal control problem (OCP) can be formulated as shown in Equation (50).

$$\min_{[U_0, U_1, \cdots, U_{N-1}]} J \qquad (50)$$

$$\text{such that } \begin{cases} X_{k+1} = g(X_k, U_k) \\ U_{min} \leq U_0, U_1, \cdots, U_{N-1} \leq U_{max} \end{cases}$$

$$\text{where } \begin{cases} U_{min} = \begin{bmatrix} V_{T,min} \\ \delta_{T,min} \end{bmatrix} \\ U_{max} = \begin{bmatrix} V_{T,max} \\ \delta_{T,max} \end{bmatrix} \end{cases}$$

The goal of this OCP is to solve for a control sequence, falling within input bounds, that minimizes the cost function while complying with the system behaviors as well.

Incorporating the discrete trailer kinematics into the OCP yields a nonlinear static optimization problem as shown in Equation (51), which can be solved using various static optimization solvers available such as [37]. The terms in Equation (51) are defined as shown in Equations (52) to (55).

$$\min_{\overline{U}} \overline{X}^T \overline{Q} \overline{X} + \overline{U}^T \overline{R} \overline{U} + X_N^T P X_N \qquad (51)$$

$$\text{such that } [U_{min}{}^T, U_{min}{}^T, \cdots]^T \leq \bar{U} \leq [U_{max}{}^T, U_{max}{}^T, \cdots]^T$$

$$\bar{X} = [X_0^T \quad X_1^T \quad \cdots \quad X_{N-1}^T]^T \in \mathbb{R}^{3N \times 1} \tag{52}$$

$$\bar{U} = [U_0^T \quad U_1^T \quad \cdots \quad U_{N-1}^T]^T \in \mathbb{R}^{2N \times 1} \tag{53}$$

$$\bar{Q} = diag(Q, Q, \cdots) \in \mathbb{R}^{3N \times 3N} \tag{54}$$

$$\bar{R} = diag(R, R, \cdots) \in \mathbb{R}^{2N \times 2N} \tag{55}$$

The NMPC approach described here is closed loop in the sense that only the first element in $\bar{U}$, which is $U_0$, will be applied to progress the system to its next time step, and the optimization routine will be run again using the updated states. It should also be remarked that since the trailer axle speed is part of the inputs to be optimized, the termination condition of the path-planning process can simply be set as when the trailer speed in the solution becomes zero, as zero speed choice indicates that additional motions cannot improve the results any further.

Once the trailer inputs (trailer axle speed and virtual steering angle at the trailer hitch) have been solved with optimization, Equations (44) and (45) can be used to calculate the actual vehicle-trailer system inputs (vehicle rear axle speed and vehicle front axle steering angle) that would allow the trailer to follow the optimized path.

### 4.2. Input Constraints

Since the control inputs generated by the NMPC routine described in Section 4.1 are virtual inputs for the trailer unit, the input constraints $U_{max}$ and $U_{min}$, which are the upper and lower bounds of trailer axle speed and virtual steering angle, should be designed with care. The trailer speed constraints can be set somewhat freely, as the vehicle is most likely able to deliver the speed required for the maneuver considering its low-speed nature. The virtual steering limits, however, must be designed such that its virtual-actual steering angle mapping under the current hitch angle will not yield an actual steering angle that is unfeasible for the vehicle front axle. At the same time, the virtual steering angle will require its own feasible value range so that it can ensure reasonable trailer orientation behaviors. This subsection hence aims to provide some design details for the virtual steering angle constraints ($\delta_{T,min}$ and $\delta_{T,max}$).

If we denote the vehicle front wheel steering angle range as $[\delta_{f,min}, \delta_{f,max}]$, then Equation (56) as shown below that maps $\delta_f$ to $\delta_T$ can be used to generate the mapped upper and lower bounds of the virtual steering angle, denoted as $[\delta_{T,lb1}, \delta_{T,ub1}]$, under the current hitch angle.

$$\delta_T = \text{atan}\left(\frac{L \cdot \sin(\Delta\psi) - L_H \cdot \cos(\Delta\psi)\tan(\delta_f)}{L \cdot \cos(\Delta\psi) + L_H \cdot \sin(\Delta\psi)\tan(\delta_f)}\right) \tag{56}$$

If we further define the upper and lower bounds of the virtual steering angle that guarantee reasonable trailer orientation behaviors as $[\delta_{T,lb2}, \delta_{T,ub2}]$, then the virtual steering input constraints ($\delta_{T,min}$ and $\delta_{T,max}$ elements in $U_{min}$ and $U_{max}$) can be defined as shown in Equation (57).

$$[\delta_{T,min}, \delta_{T,max}] = [\delta_{T,lb1}, \delta_{T,ub1}] \cap [\delta_{T,lb2}, \delta_{T,ub2}] \quad (57)$$

It should be noted that this virtual steering constraint setting can only guarantee that the mapped actual steering angle does not exceed its limits at the current time step. As model prediction proceeds into future steps within the horizon, the hitch angle of the vehicle-trailer system is expected to change, hence potentially rendering the virtual steering limits invalid. However, since only the first step of the optimized input sequence (corresponding to current time step) will be applied, this setup can ensure that the actual vehicle steering angle always stays within its limits.

A simple numeric example is presented here to demonstrate the proposed input constraints design. Model parameter values listed in Table 2 are used. If one defines the current hitch angle of the vehicle-trailer system ($\Delta\psi$) and the vehicle front wheel steering angle range ($\delta_{f,min}$ and $\delta_{f,max}$) to take the values listed in Table 3, then applying Equation (56) will yield $[\delta_{T,lb1}, \delta_{T,ub1}] = [-2.2512, 32.2512] \, [deg]$. If one continues to define a reasonable set of trailer virtual steering angle limits ($\delta_{T,lb2}$ and $\delta_{T,ub2}$) as shown in Table 3, one can apply Equation (57) to narrow down the range of admissible virtual steering to $[\delta_{T,min}, \delta_{T,max}] = [-2.2512, 28.6479] \, [deg]$.

**Table 3.** Parameter value choices for input constraints design numeric example

| Parameter | Value Choice |
|---|---|
| $\Delta\psi$ | 0.2618 [rad] = 15 [deg] |
| $[\delta_{f,min}, \delta_{f,max}]$ | [-0.75, 0.75] [rad] = [-42.9718, 42.9718] [deg] |
| $[\delta_{T,lb2}, \delta_{T,ub2}]$ | [-0.5, 0.5] [rad] = [-28.6479, 28.6479] [deg] |

## 5. Forward Repositioning

In the occasion that backward motions of the vehicle-trailer combination alone cannot satisfactorily position the trailer into the parking space with proper orientation or the final hitch angle of the system is too large, pulling forward and backing up again will be required. The forward repositioning maneuver has two main purposes: 1) obtain a more favorable system hitch angle so that it is easier to orientate the trailer unit as desired during further reverse motions; 2) open up some space for the system so that further backup maneuvers have room for adjustments. While any additional backup maneuvers are to be handled with the proposed NMPC formulation detailed in Section 4, forward repositioning motion is handled with a pure-pursuit path-tracking controller that only applies to the vehicle.

The forward path is designed as a straight line that extends out of the parking space with its orientation aligning with the desired final trailer orientation. Figure 6 shows the schematic of such a forward path. Once the path is defined, the pure-pursuit controller can be designed as explained in detail in [44]. It should be remarked that the pure-pursuit controller here is designed to be applied to the vehicle unit in forward motions only, so the complications involving the trailer unit can be disregarded during the design process. Nevertheless, we do need to terminate the vehicle-trailer forward motion if both of the following conditions are met: 1) the hitch angle is smaller than a pre-defined angle threshold; 2) the vehicle-trailer system is further away from the parking space than a pre-defined distance threshold.

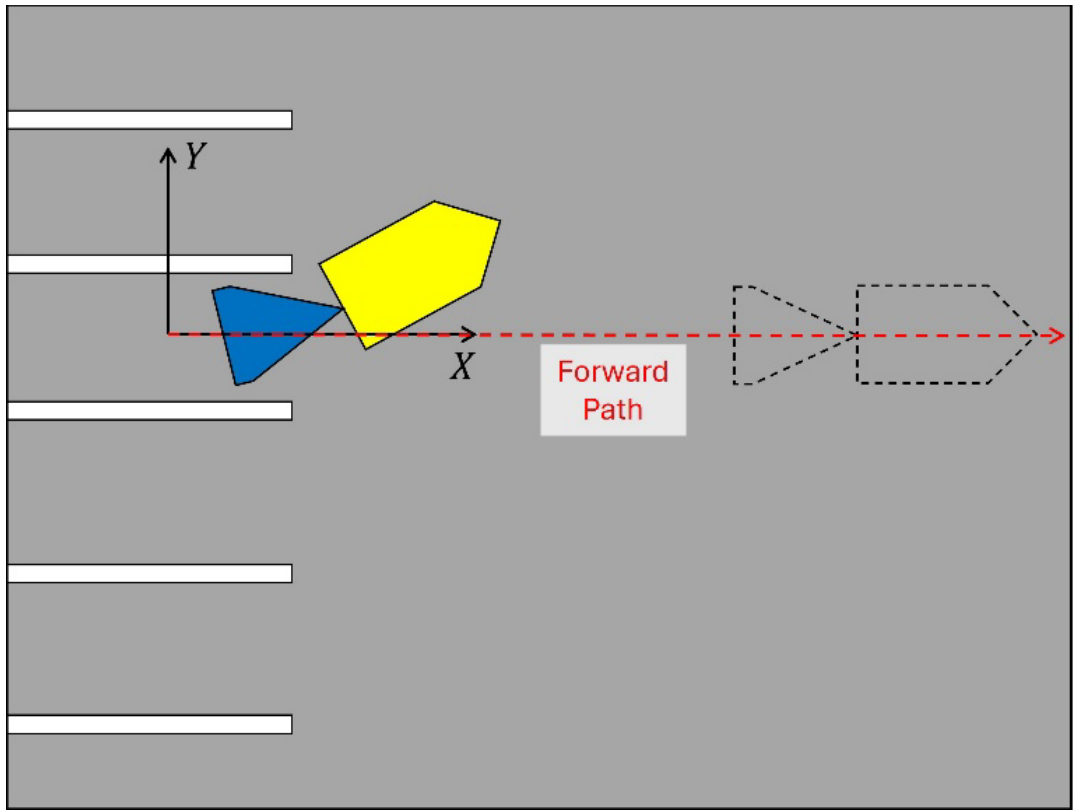

**Figure 6.** Forward path for forward repositioning

Once the forward repositioning maneuver is completed, the NMPC-based backup action will be applied again to reverse the vehicle-trailer system into the desired parking space, hence finishing the parking maneuver. The overall flowchart of this 3-staged path-planning process is displayed in Figure 7.

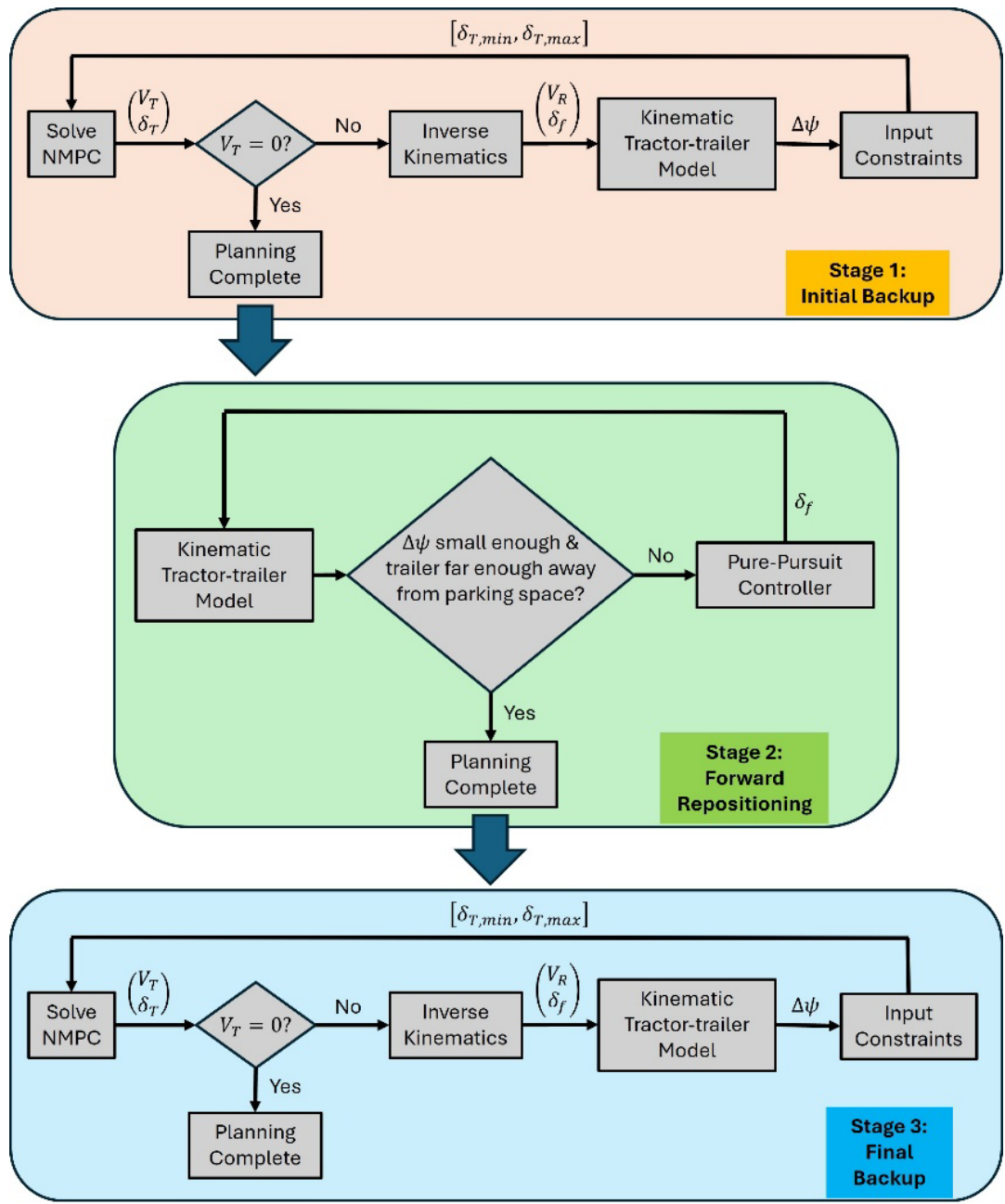

**Figure 7.** Overall path-planning process flowchart

## 6. Implementation Results

### 6.1. Simulation Case Study

      Simulation studies are first conducted to test the efficacy of the proposed method. The value choices for various settings are listed in Table 4. Overall, four tests are conducted, each with different initial conditions, to demonstrate the ability of the proposed

path-planning and control routine to adapt to numerous scenarios and system configurations.

**Table 4.** Simulation case study value choices

| Model Parameter | Value Choice |
|---|---|
| $L$ | 2.896 [m] |
| $L_H$ | 1.159 [m] (passenger vehicle) |
| $L_T$ | 2.693 [m] |
| $d_t$ | 0.1 [sec] |
| $N$ | 10 |
| $Q, P$ | $\begin{bmatrix} 1 & 0 & 0 \\ 0 & 10 & 0 \\ 0 & 0 & 10 \end{bmatrix}$ |
| $R$ | $\begin{bmatrix} 0 & 0 \\ 0 & 0.1 \end{bmatrix}$ |
| $[V_{T,min}, V_{T,max}]$ | [-1, 0] [m/sec] |
| $[\delta_{f,min}, \delta_{f,max}]$ | [-0.75, 0.75] [rad] = [-42.9718, 42.9718] [deg] |
| $[\delta_{T,lb2}, \delta_{T,ub2}]$ | [-0.5, 0.5] [rad] = [-28.6479, 28.6479] [deg] |

Scenario 1 aims to replicate a standard parallel parking maneuver. The results of the simulation are presented in Figure 8 and Figure 9. The maneuver is conducted in three stages. Stage 1 motion is performed with NMPC-based path-planning routine, and it can be observed that the trailer unit can be guided close to the desired parking position, with cost function value continuously decreasing during the process. It can also be observed that the algorithm chooses the highest trailer axle reverse speed for as long as possible, as the goal of each optimization loop is to get as close to the final position and orientation as possible. Stage 1 motion is concluded with the choice of zero trailer axle speed, as the NMPC routine will select zero speed when further motions can no longer reduce cost function value. Stage 2 is the forward repositioning maneuver completed by the pure-pursuit controller applied to the vehicle unit. It can be observed that the forward motion allows the vehicle-trailer system to obtain a much more favorable configuration and opens the space for further backward adjustments of the system. Stage 3, identical to Stage 1, is a reverse maneuver performed with the same NMPC routine. It can be observed that thanks to the forward motion in Stage 2, this stage requires much less trajectory adjustments compared to Stage 1. An additional note is that the tractor vehicle steering commands remain within the specified range for the entire duration of the parking maneuver, proving the effectiveness of the input constraint design.

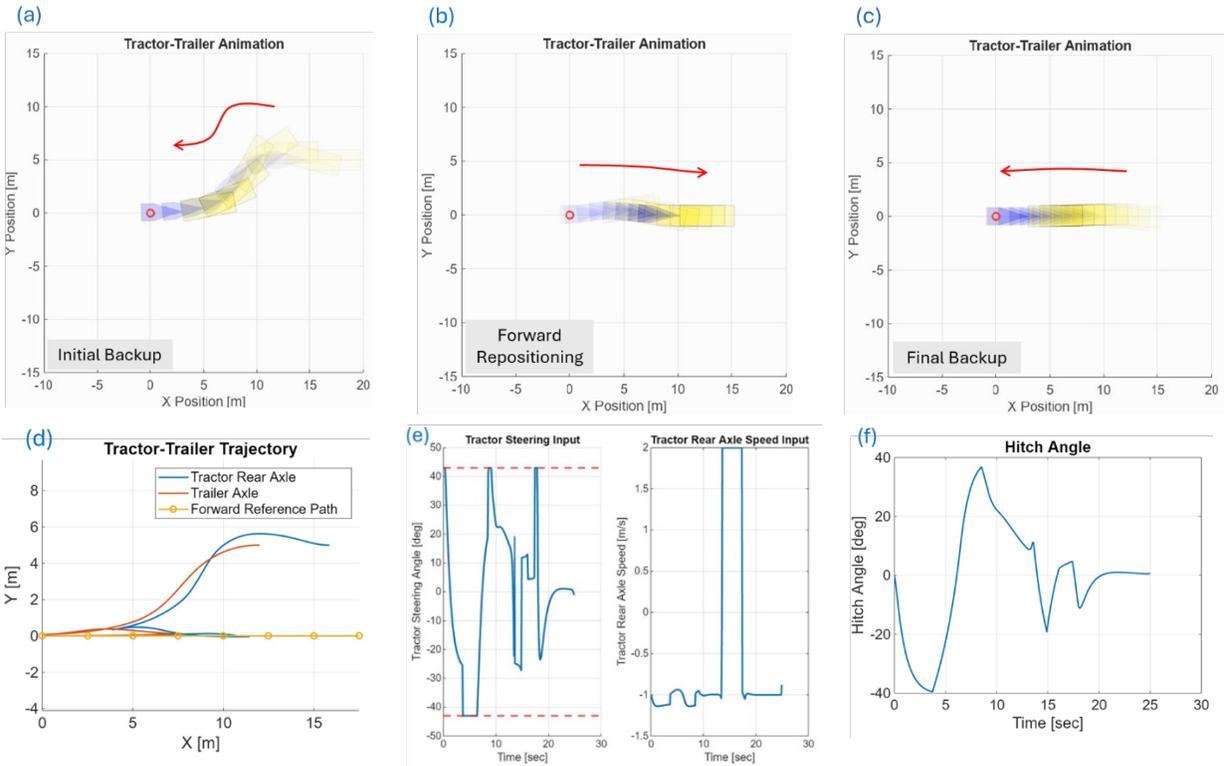

**Figure 8.** Simulation results for scenario 1: (a) Stage 1 motion; (b) Stage 2 motion; (c) Stage 3 motion; (d) Overall trajectory; (e) Tractor vehicle input history; (f) Hitch angle history

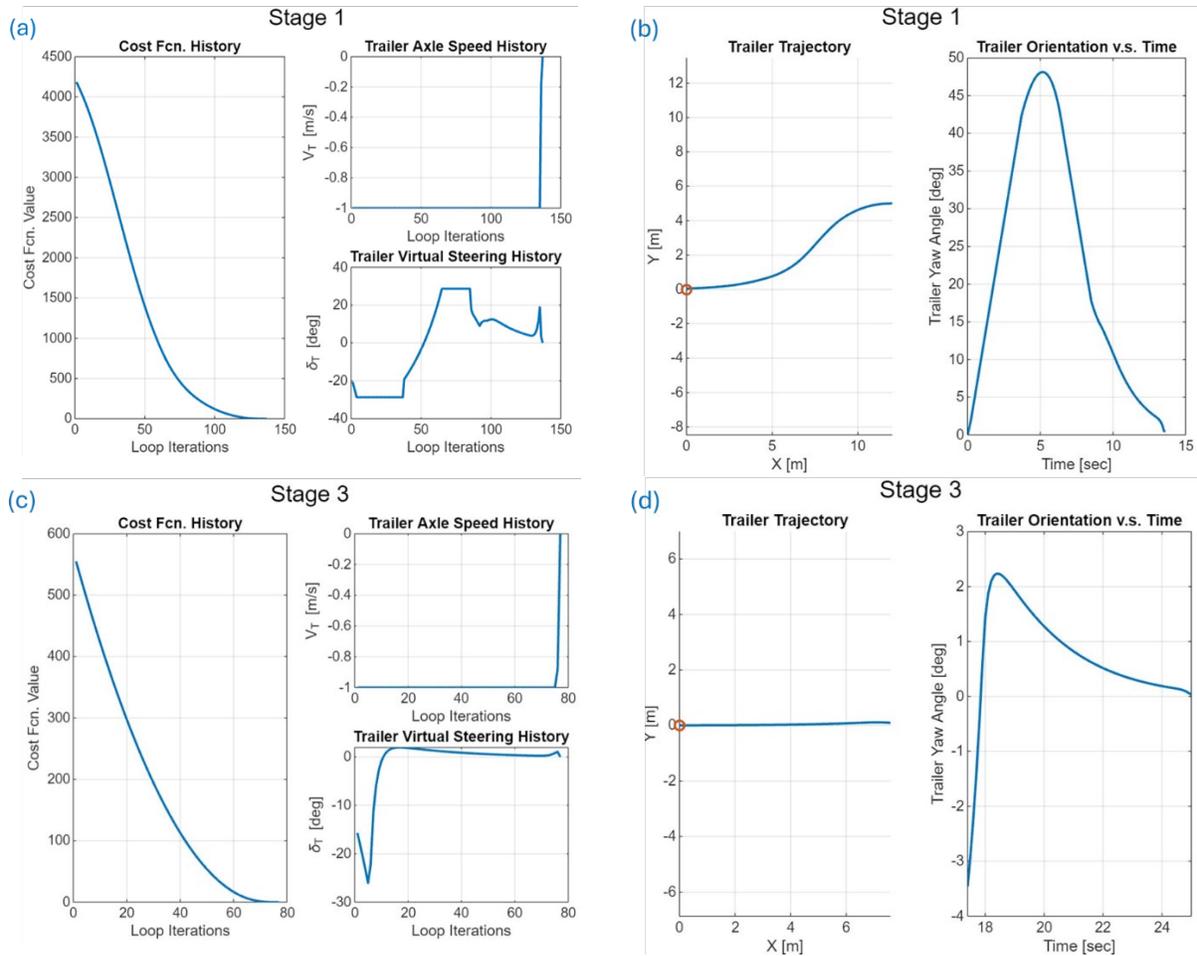

**Figure 9.** Simulation results for scenario 1: (a) Stage 1 cost function history and virtual inputs history; (b) Stage 1 trailer trajectory and trailer orientation history (c) Stage 3 cost function history and virtual inputs history; (d) Stage 3 trailer trajectory and trailer orientation history

Scenario 2 aims to replicate a parallel parking maneuver with a nonzero initial hitch angle. The results of the simulation are presented in Figure 10 and Figure 11. The maneuver is again conducted in three stages, where stage 1 and stage 3 are handled with the NMPC reverse path-planning algorithm and stage 2 is tackled with the forward pure-pursuit controller. It can be observed that since the initial configuration is not as favorable as compared to that in scenario 1, stage 1 motion cannot put the vehicle-trailer system into the parking space with an ideal orientation. As a result, stage 2 motion demonstrates more aggressive steering adjustments to achieve a better configuration. Once stage 2 has concluded, however, stage 3 is able to drive the system into the parking space much more easily. Again, it should be noted that vehicle unit steering inputs remain within the specified range for the entire duration of the maneuver.

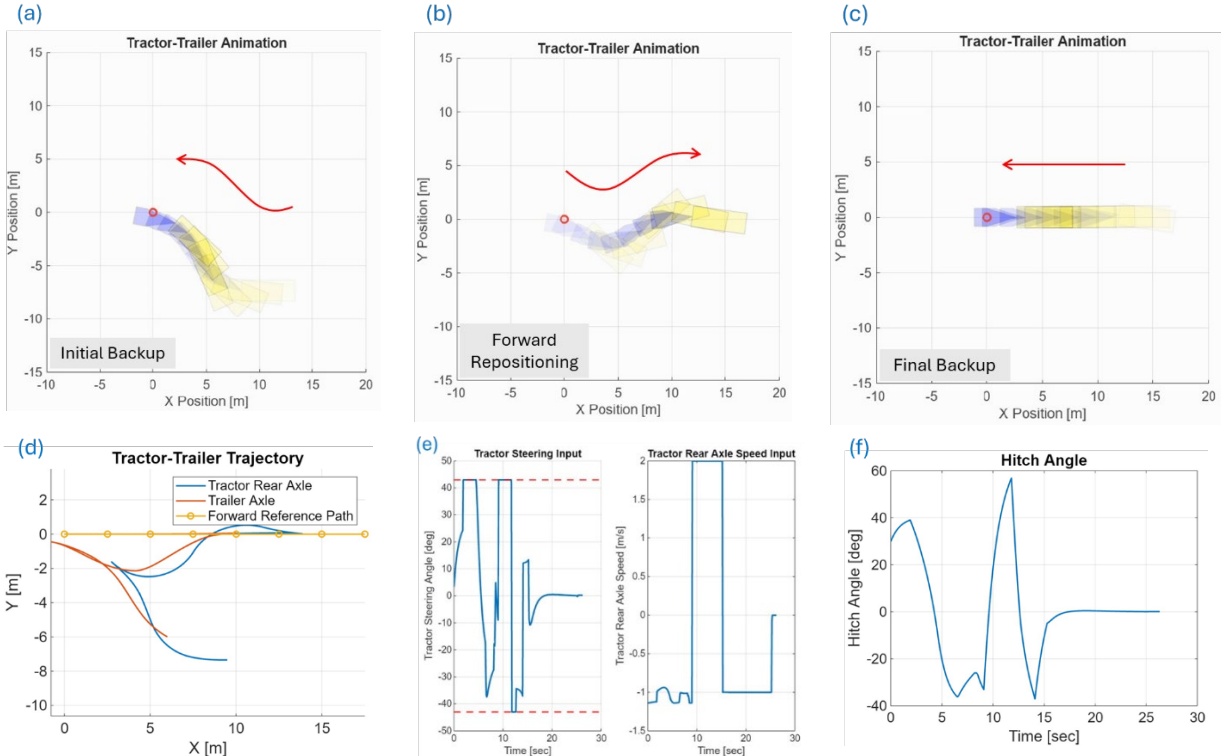

**Figure 10.** Simulation results for scenario 2: (a) Stage 1 motion; (b) Stage 2 motion; (c) Stage 3 motion; (d) Overall trajectory; (e) Tractor vehicle input history; (f) Hitch angle history

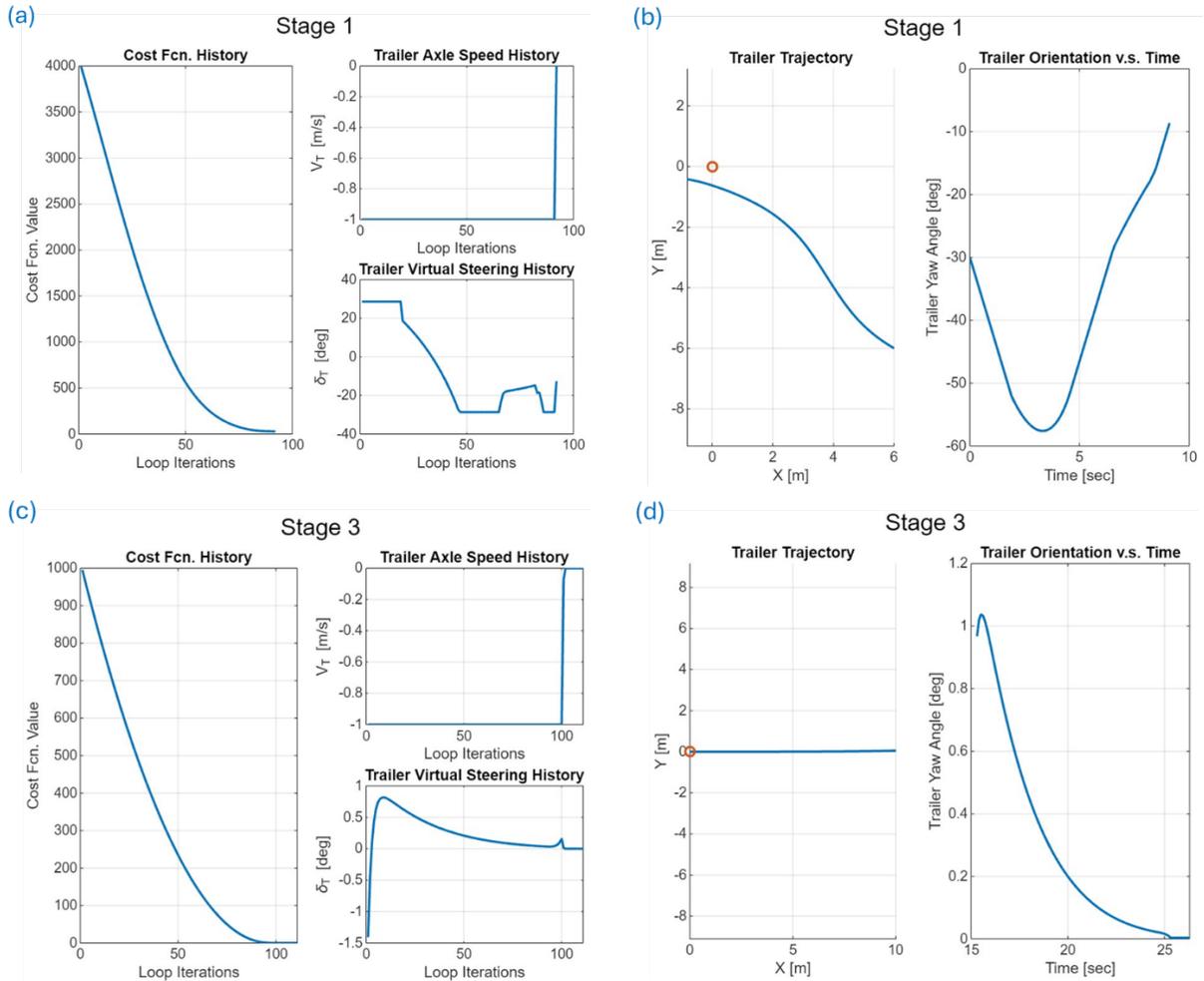

**Figure 11.** Simulation results for scenario 2: (a) Stage 1 cost function history and virtual inputs history; (b) Stage 1 trailer trajectory and trailer orientation history (c) Stage 3 cost function history and virtual inputs history; (d) Stage 3 trailer trajectory and trailer orientation history

Scenario 3 aims to replicate a standard perpendicular parking maneuver. Figure 12 and Figure 13 illustrate the results of the simulation. It can again be observed that the three-staged motion is able to achieve favorable vehicle-trailer system terminal states while keeping vehicle steering within the specified limits.

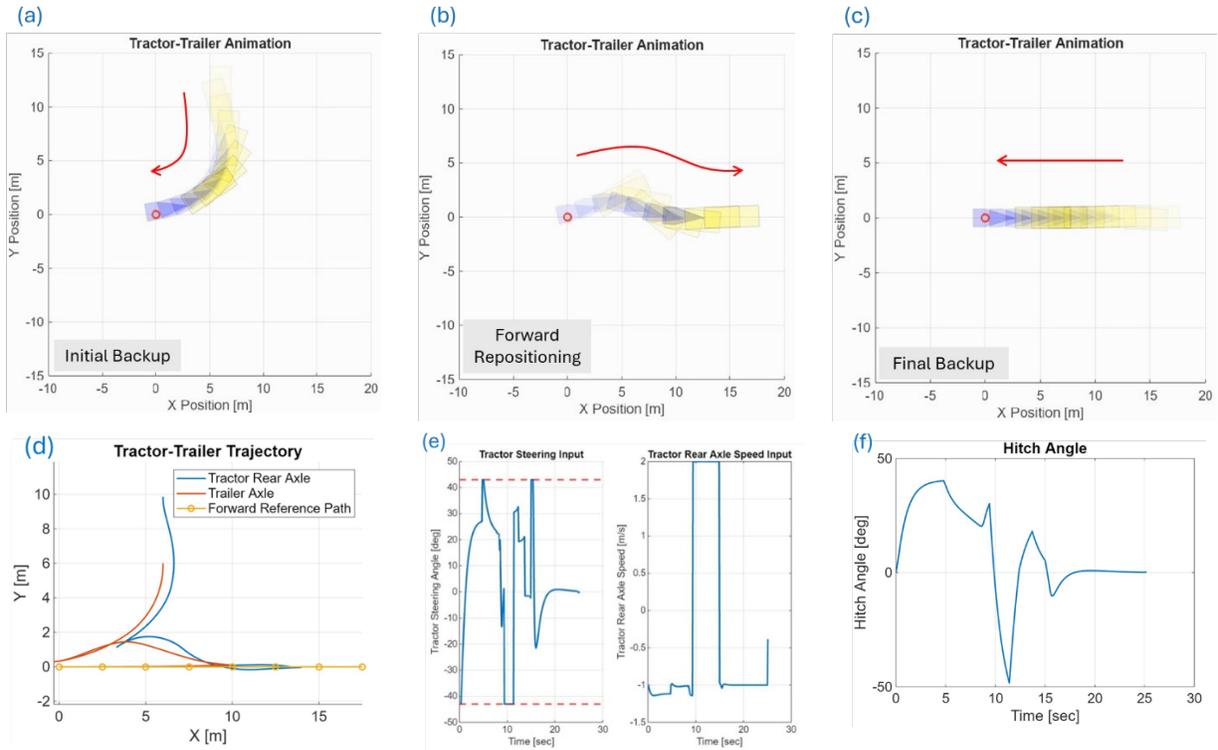

**Figure 12.** Simulation results for scenario 3: (a) Stage 1 motion; (b) Stage 2 motion; (c) Stage 3 motion; (d) Overall trajectory; (e) Tractor vehicle input history; (f) Hitch angle history

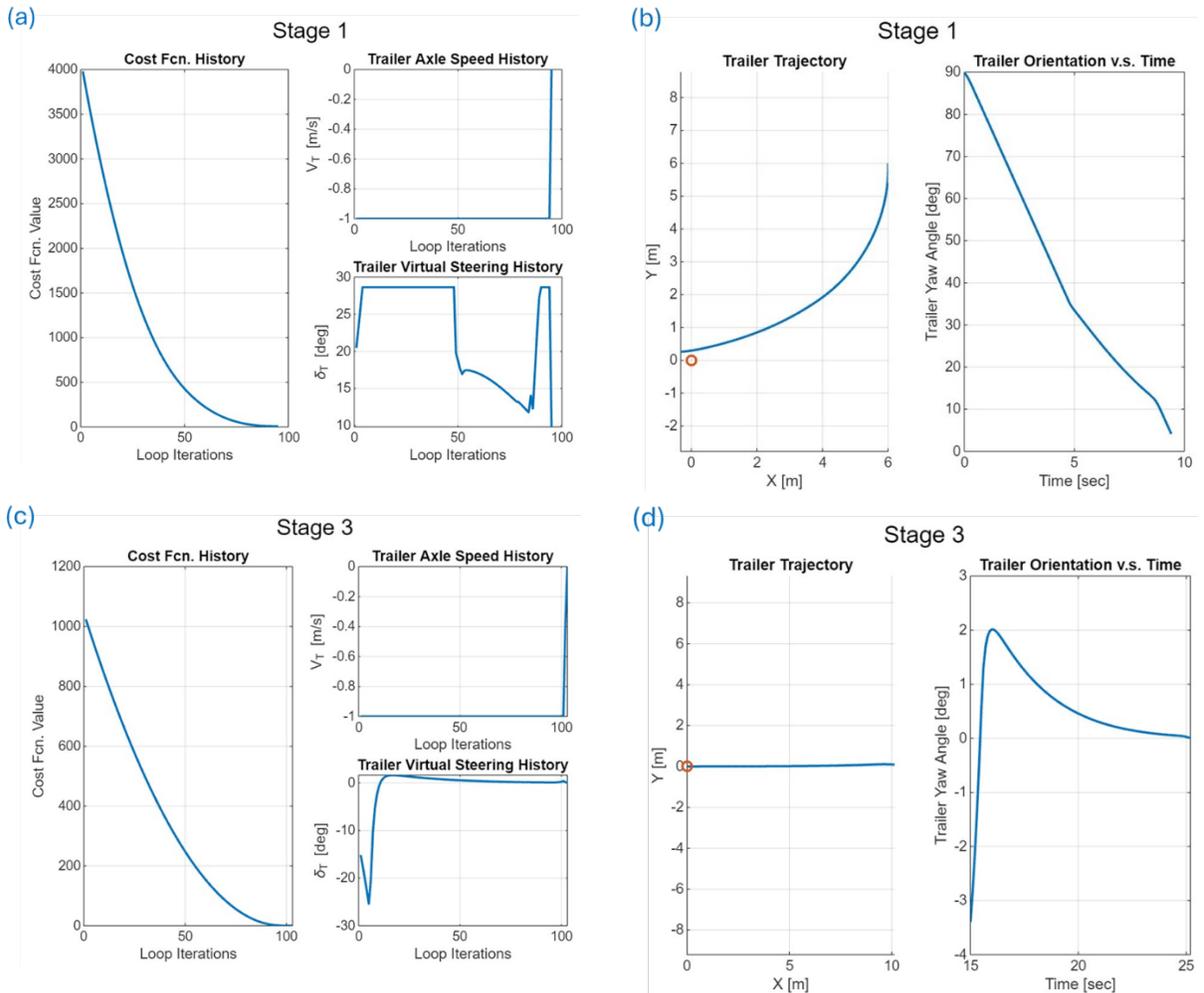

**Figure 13.** Simulation results for scenario 3: (a) Stage 1 cost function history and virtual inputs history; (b) Stage 1 trailer trajectory and trailer orientation history (c) Stage 3 cost function history and virtual inputs history; (d) Stage 3 trailer trajectory and trailer orientation history

Scenario 4 aims to recreate a generic parking maneuver that starts with a random initial position and a random nonzero initial hitch angle. Figure 14 and Figure 15 show the results of the simulation. It can be observed that yet again the proposed approach can complete the parking maneuver with the three-stage framework while keeping system input within limits.

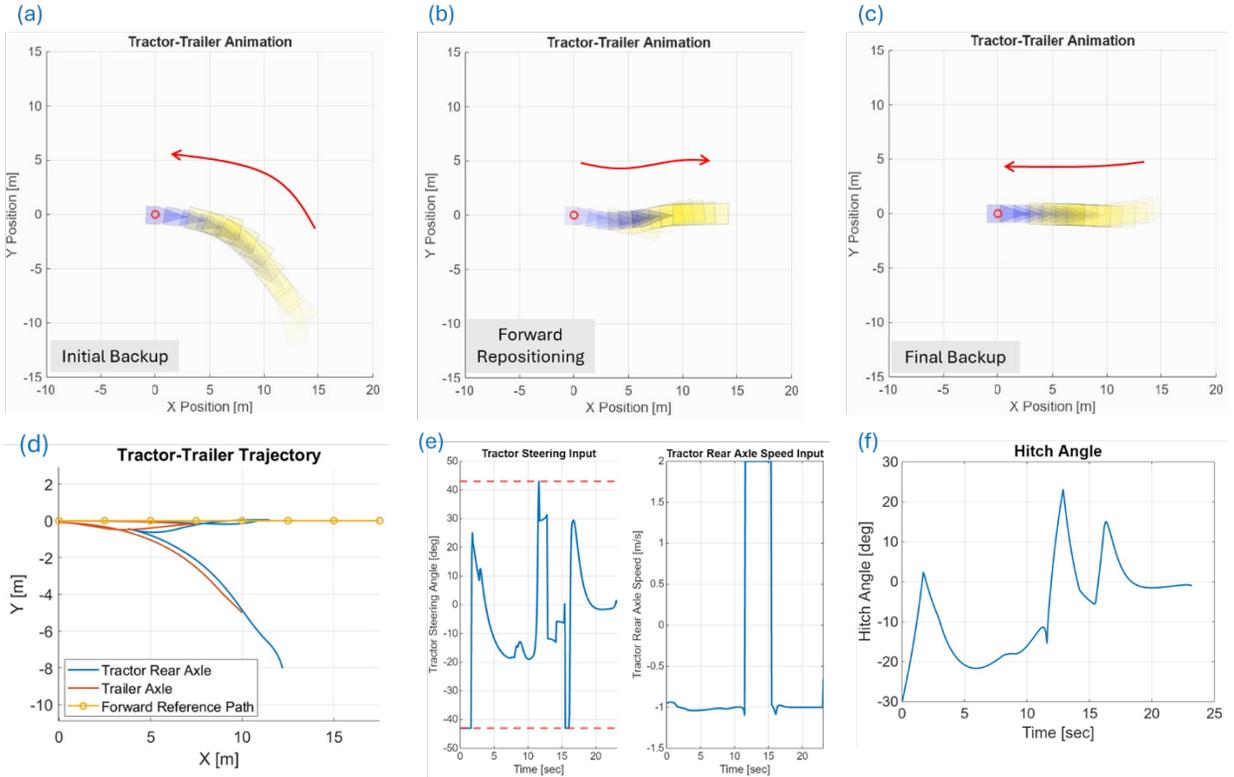

**Figure 14.** Simulation results for scenario 4: (a) Stage 1 motion; (b) Stage 2 motion; (c) Stage 3 motion; (d) Overall trajectory; (e) Tractor vehicle input history; (f) Hitch angle history

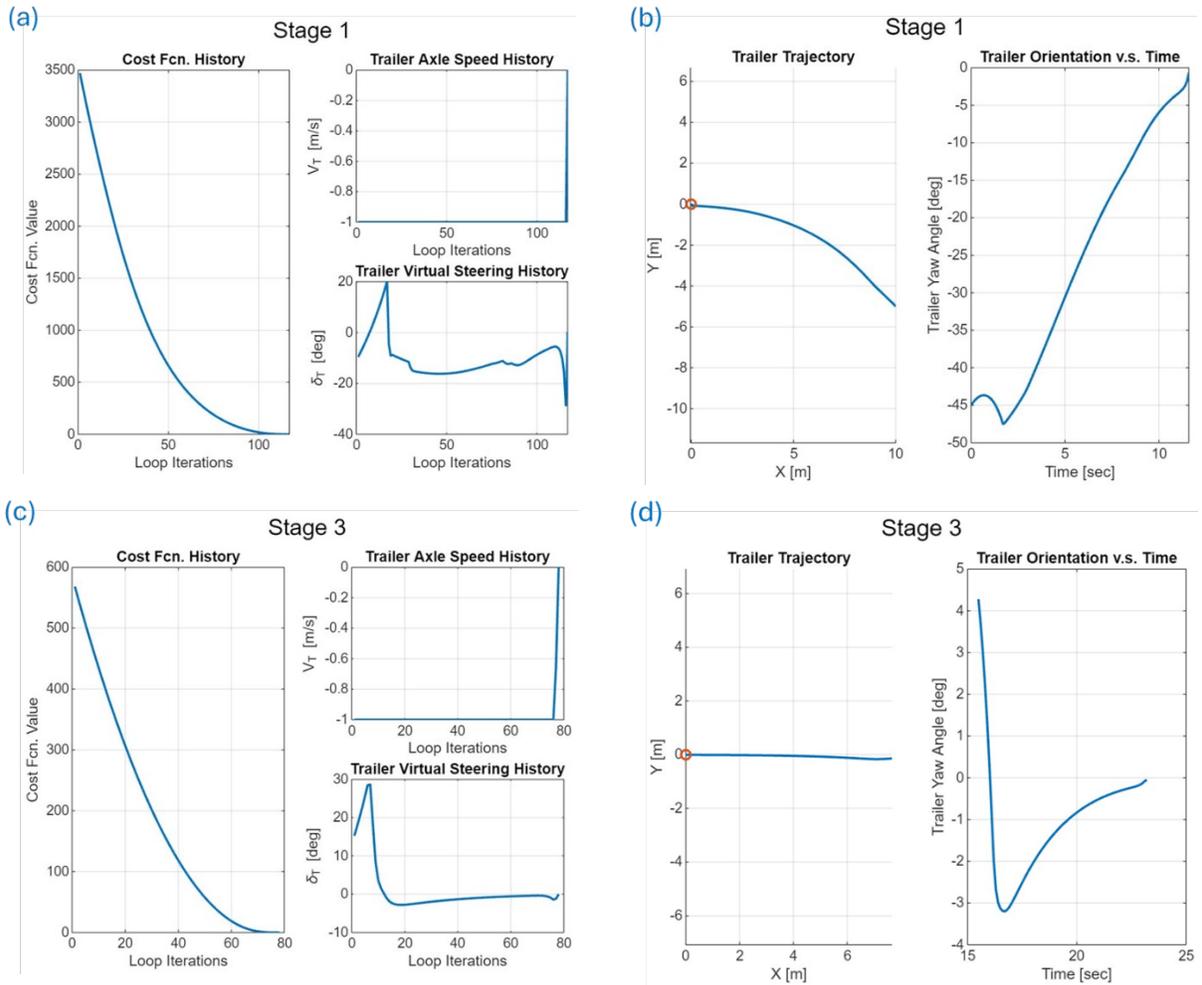

**Figure 15.** Simulation results for scenario 4: (a) Stage 1 cost function history and virtual inputs history; (b) Stage 1 trailer trajectory and trailer orientation history (c) Stage 3 cost function history and virtual inputs history; (d) Stage 3 trailer trajectory and trailer orientation history

Additional observations can be made according to the numeric results recorded in Table 5. It can be observed that across all the scenarios, the addition of stage 2 forward motions allows the vehicle-trailer system to obtain a much more favorable configuration, hence enabling stage 3 backward motions to achieve much smaller trailer distance errors and trailer orientation errors compared to stage 1 backward motions. Furthermore, the terminal hitch angles at the end of stage 3 motions are significantly smaller than those at the end of stage 1 motions for all scenarios. Despite it not being one of the design objectives of the algorithm, a small terminal hitch angle is generally desirable for a vehicle-trailer system parking maneuver.

**Table 5.** Simulation case study numeric results

|  | Stage 1 | | | Stage 3 | | |
|---|---|---|---|---|---|---|
|  | Trailer Distance Error [m] | Trailer Orientation Error [deg] | Final Hitch Angle [deg] | Trailer Distance Error [m] | Trailer Orientation Error [deg] | Final Hitch Angle [deg] |
| Scenario 1 | 0.0525 | 0.4976 | 11.0584 | 0.0033 | 0.0319 | 0.6667 |
| Scenario 2 | 0.9201 | -8.6355 | -33.0678 | 4.4619e-4 | 0.0044 | 0.0964 |
| Scenario 3 | 0.4134 | 4.1443 | 30.2587 | 0.0011 | 0.0114 | 0.2324 |
| Scenario 4 | 0.0846 | -0.6137 | -15.2986 | 0.0049 | -0.0495 | -0.9789 |

## 6.2. Hardware-in-the-loop (HIL) Testing

With simulation studies in Section 6.1 demonstrating satisfactory performance of the proposed approach, hardware-in-the-loop (HIL) testing is carried out to further test the suitability of this method for real vehicle applications. In HIL experiments, physical hardware that would be used in real vehicle experiments are integrated into the simulator to make the simulation test more realistic. Converting the simulation to run online is needed to achieve HIL functionality. Figure 16(a) shows the HIL simulator architecture and information flowchart, while Figure 16(b) displays the physical layout of the simulator used in this analysis.

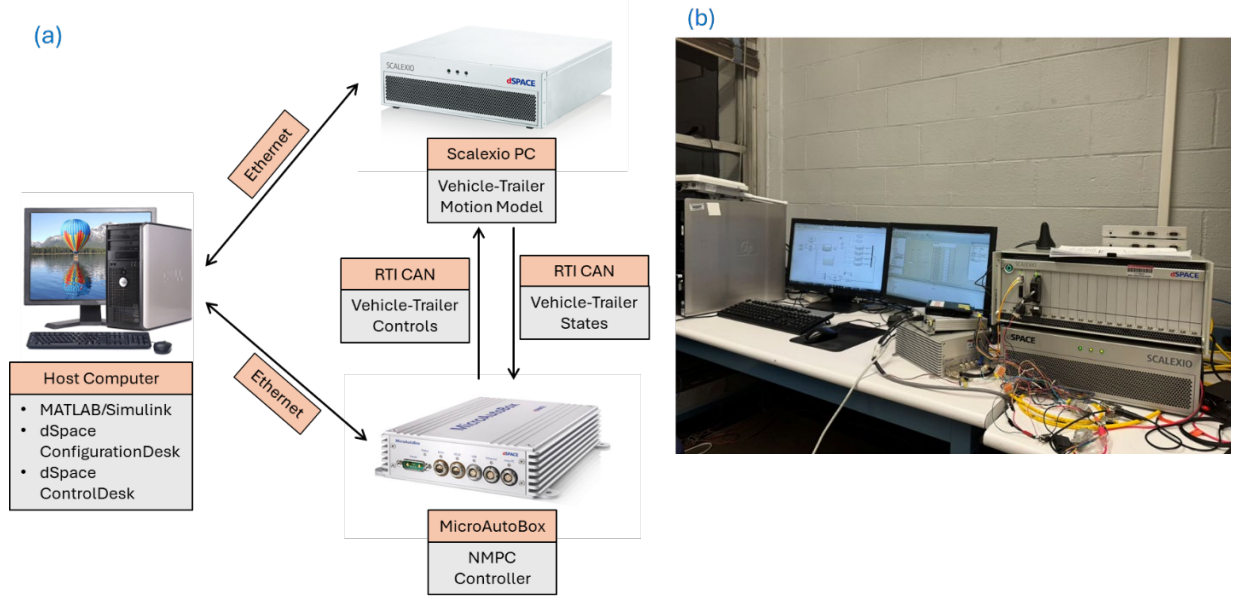

**Figure 16.** (a) HIL architecture and information flowchart; (b) HIL simulator layout

The results of the HIL test are displayed in Figure 17. The model parameters used in this test are consistent with the value choices listed in Table 4. A slightly more unfavorable system initial condition compared to scenario 1 is used in this case. It can be observed that the proposed NMPC-based routine is able to use tractor steering inputs that are within the limits to direct the vehicle-trailer system to the desired terminal position and orientation despite the less than desirable initial pose. The routine also proves itself to be adequate for online operations with its small dimension enabled by the inverse kinematics calculation, as its calculation can be easily carried out in the MicroAutoBox unit without triggering task overrun.

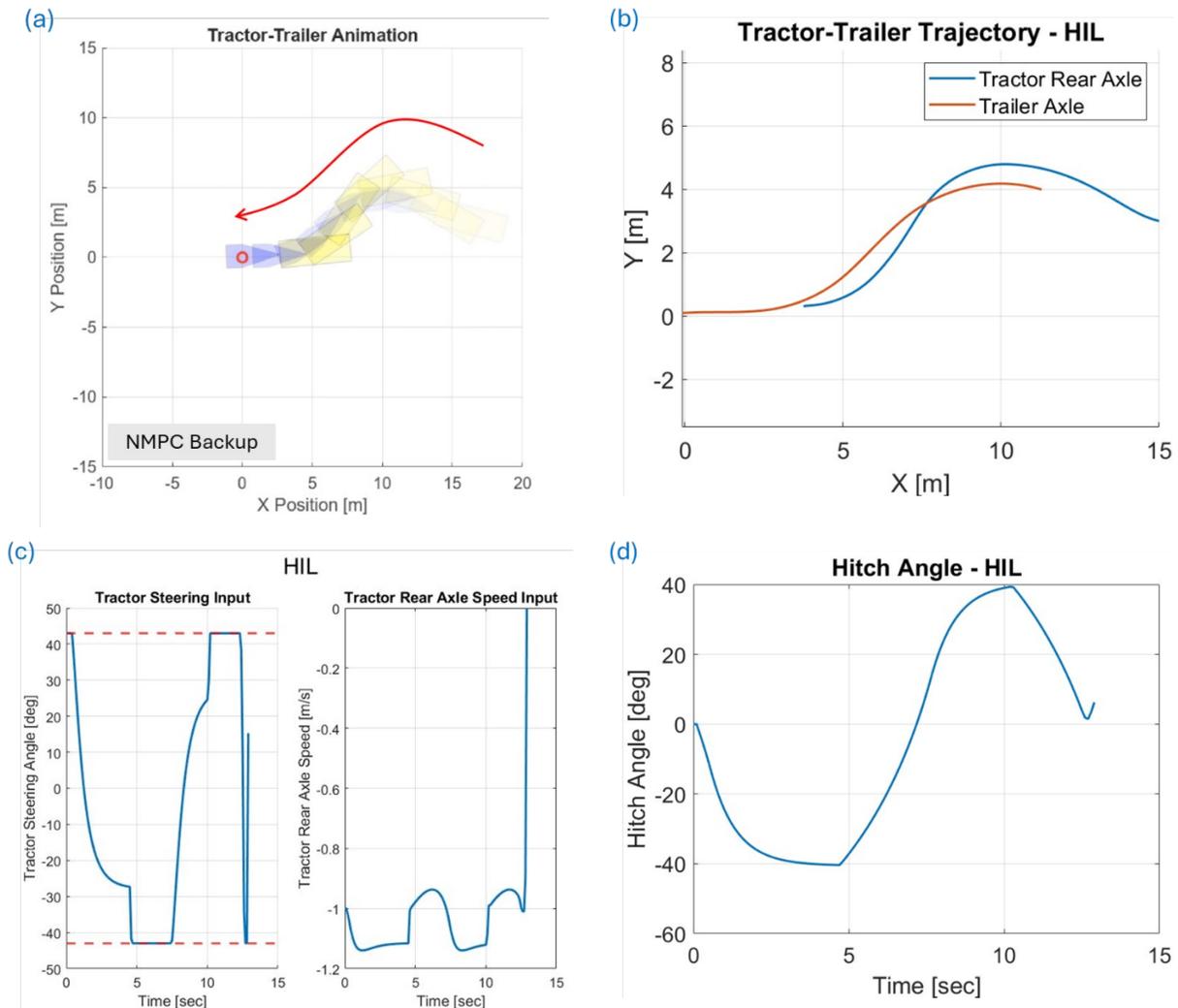

**Figure 17.** HIL results: (a) NMPC reverse motion; (b) Vehicle-trailer trajectory; (e) Tractor vehicle input history; (f) Hitch angle history

## 7. Conclusion and Future Work

This paper proposed an optimization-based path-planning and path-tracking strategy to tackle the difficult problem of vehicle with trailer system reverse parking. A generic kinematic vehicle-trailer model with one trailer configuration was first derived. The inverse kinematics of the system was also derived to allow for the trailer being regarded as a standalone unit in the path-planning process, where 'virtual' inputs at the trailer can later be propagated to the actual inputs at the tractor vehicle. A nonlinear model predictive control (NMPC) based optimal control problem was formulated to plan a kinematically feasible trajectory for the trailer unit, and optional forward repositioning maneuver achieved with tractor vehicle forward pure-pursuit controller was added as well. Simulation studies and HIL testing demonstrated the effectiveness of the proposed routine with its small dimension and ease of implementation in real-time operations. For future work, obstacle avoidance features can be integrated into the NMPC formulation, either as hard constraints or soft constraints, to further expand the scope of application for the proposed approach. Alternative variants of the NMPC formulation can also be attempted. Examples include the following: 1) separate prediction horizon and control horizon; 2) varying cost function weightings; 3) generating coarse initial paths with searching methods such as Hybrid A* and rapidly exploring random trees (RRT) and designing the cost function of the OCP to mimic the initial paths.

**References**

[1] R. Hussain and S. Zeadally, "Autonomous Cars: Research Results, Issues, and Future Challenges," *IEEE Commun. Surv. Tutor.*, vol. 21, no. 2, pp. 1275–1313, 2019, doi: 10.1109/COMST.2018.2869360.

[2] S. Y. Gelbal, B. A. Guvenc, and L. Guvenc, "SmartShuttle: a unified, scalable and replicable approach to connected and automated driving in a smart city," in *Proceedings of the 2nd International Workshop on Science of Smart City Operations and Platforms Engineering*, in SCOPE '17. New York, NY, USA: Association for Computing Machinery, Apr. 2017, pp. 57–62. doi: 10.1145/3063386.3063761.

[3] B. Wen, S. Y. Gelbal, B. A. Guvenc, and L. Guvenc, "Localization and Perception for Control and Decision-Making of a Low-Speed Autonomous Shuttle in a Campus Pilot Deployment," *SAE Int. J. Connect. Autom. Veh.*, vol. 1, no. 2, Art. no. 12-01-02–0003, Nov. 2018, doi: 10.4271/12-01-02-0003.

[4] S. Velupillai and L. Guvenc, "Laser Scanners for Driver-Assistance Systems in Intelligent Vehicles [Applications of Control]," in *IEEE Control Systems Magazine*, vol. 29, no. 2, pp. 17-19, April 2009, doi: 10.1109/MSP.2008.931716.

[5] Kavas-Torris, O., Gelbal, S. Y., Cantas, M. R., Aksun Guvenc, B., and Guvenc, L., "V2X Communication between Connected and Automated Vehicles (CAVs) and Unmanned Aerial Vehicles (UAVs)," *Sensors*, 2022, 22(22), 8941. https://doi.org/10.3390/s22228941.

[6] S. Y. Gelbal, S. Arslan, H. Wang, B. Aksun-Guvenc and L. Guvenc, "Elastic band based pedestrian collision avoidance using V2X communication," 2017 *IEEE Intelligent Vehicles Symposium* (IV), Los Angeles, CA, USA, 2017, pp. 270-276, doi: 10.1109/IVS.2017.7995731.

[7] O. Ararat, B. Aksun-Guvenc, "Development of a Collision Avoidance Algorithm Using Elastic Band Theory," *IFAC Proceedings Volumes*, Volume 41, Issue 2, 2008, Pages 8520-8525, https://doi.org/10.3182/20080706-5-KR-1001.01440.

[8] Ma, F., Wang, J., Yang, Y. et al, "Stability Design for the Homogeneous Platoon with Communication Time Delay," Automotive Innovation, 3, pp. 101–110 (2020). https://doi.org/10.1007/s42154-020-00102-4.

[9] Gelbal, S.Y., Aksun-Guvenc, B., Guvenc, L. "Collision Avoidance of Low Speed Autonomous Shuttles with Pedestrians," I*nternational Journal of Automotive Technology,* 21, 903–917 (2020). https://doi.org/10.1007/s12239-020-0087-7

[10] L. Claussmann, M. Revilloud, D. Gruyer, and S. Glaser, "A Review of Motion Planning for Highway Autonomous Driving," *IEEE Trans. Intell. Transp. Syst.*, vol. 21, no. 5, pp. 1826–1848, May 2020, doi: 10.1109/TITS.2019.2913998.

[11] L. Guvenc, B. A. Guvenc, and M. T. Emirler, "Connected and Autonomous Vehicles," in *Internet of Things and Data Analytics Handbook*, John Wiley & Sons, Ltd, 2017, pp. 581–595. doi: 10.1002/9781119173601.ch35.

[12] X. Cao *et al.*, "Development of an Advisory System for Parking a Car with Trailer," SAE International, Warrendale, PA, SAE Technical Paper 2025-01–8035, Apr. 2025. doi: 10.4271/2025-01-8035.

[13] F. Jindra, "Handling Characteristics of Tractor-Trailer Combinations," *SAE Trans.*, vol. 74, pp. 378–394, 1966.

[14] L. Alexander, M. Donath, M. Hennessey, V. Morellas, and C. Shankwitz, "A Lateral Dynamic Model of a Tractor-Trailer: Experimental Validation," University of Minnesota Twin Cities, Minneapolis, MN, USA, Report Number 97-18, Nov. 1996.

[15] T. Lei, J. Wang, and Z. Yao, "Modelling and Stability Analysis of Articulated Vehicles," *Appl. Sci.*, vol. 11, no. 8, Art. no. 8, Jan. 2021, doi: 10.3390/app11083663.

[16] Y. Gao, Y. Shen, T. Xu, W. Zhang, and L. Guvenc, "Oscillatory Yaw Motion Control for Hydraulic Power Steering Articulated Vehicles Considering the Influence of Varying Bulk Modulus," *IEEE Trans. Control Syst. Technol.*, vol. 27, no. 3, pp. 1284–1292, May 2019, doi: 10.1109/TCST.2018.2803746.

[17] M. Luijten, "Lateral Dynamic Behaviour of Articulated Commercial Vehicles," Thesis, Mechanical Engineering, Eindhoven University of Technology, Eindhoven, Netherlands, 2010.

[18] S. M. Wolfe, "Heavy Truck Modeling and Estimation for Vehicle-to-Vehicle Collision Avoidance Systems," Dissertation, Mechanical Engineering, The Ohio State University, Columbus, OH, USA, 2014.

[19] B. Jia, H. Ren, Z. Yang, Z. Lou, and X. He, "Lateral Dynamic Model for Full-Size Semi-Trailer Container Trucks with Practical Evaluation," in *2023 35th Chinese Control and Decision Conference (CCDC)*, May 2023, pp. 4295–4300. doi: 10.1109/CCDC58219.2023.10327486.

[20] Z. Brock, J. Nelson, and R. L. Hatton, "A Comparison of Lateral Dynamic Models for Tractor-Trailer Systems," in *2019 IEEE Intelligent Vehicles Symposium (IV)*, Jun. 2019, pp. 2052–2059. doi: 10.1109/IVS.2019.8814286.

[21] Y. He and J. Ren, "A Comparative Study of Car-Trailer Dynamics Models," *SAE Int. J. Passeng. Cars - Mech. Syst.*, vol. 6, no. 1, Art. no. 2013-01–0695, Apr. 2013, doi: 10.4271/2013-01-0695.

[22] M. Werling, P. Reinisch, M. Heidingsfeld, and K. Gresser, "Reversing the General One-Trailer System: Asymptotic Curvature Stabilization and Path Tracking," *IEEE Trans. Intell. Transp. Syst.*, vol. 15, no. 2, pp. 627–636, Apr. 2014, doi: 10.1109/TITS.2013.2285602.

[23] R. Kusumakar, K. Kural, A. Tomar, and B. Pyman, "Autonomous Parking for Articulated Vehicles," Thesis, HAN University of Applied Science, Nijmegen, Netherlands, 2017. doi: 10.13140/RG.2.2.30533.76009.

[24] Adityen Sudhakaran, "Autonomous Parallel Parking of a Scaled Tractor Semi-Trailer," Thesis, Mechanical Engineering, Eindhoven University of Technology, Eindhoven, Netherlands, 2018.

[25] P. Rouchon, M. Fliess, J. Levine, and P. Martin, "Flatness, motion planning and trailer systems," in *Proceedings of 32nd IEEE Conference on Decision and Control*, Dec. 1993, pp. 2700–2705 vol.3. doi: 10.1109/CDC.1993.325686.

[26] C. Pradalier and K. Usher, "A simple and efficient control scheme to reverse a tractor-trailer system on a trajectory," in *Proceedings 2007 IEEE International Conference on Robotics and Automation*, Apr. 2007, pp. 2208–2214. doi: 10.1109/ROBOT.2007.363648.

[27] C. Pradalier and K. Usher, "Experiments in Autonomous Reversing of a Tractor-Trailer System," in *Field and Service Robotics: Results of the 6th International Conference*, C. Laugier and R. Siegwart, Eds., Berlin, Heidelberg: Springer, 2008, pp. 475–484. doi: 10.1007/978-3-540-75404-6_45.

[28] Z. Leng and M. Minor, "A simple tractor-trailer backing control law for path following," in *2010 IEEE/RSJ International Conference on Intelligent Robots and Systems*, Oct. 2010, pp. 5538–5542. doi: 10.1109/IROS.2010.5650489.

[29] Z. Li, H. Cheng, J. Ma, and H. Zhou, "Research on Parking Control of Semi-trailer Truck," in *2020 4th CAA International Conference on Vehicular Control and Intelligence (CVCI)*, Dec. 2020, pp. 424–429. doi: 10.1109/CVCI51460.2020.9338617.

[30] A. Tomar *et al.*, "Path following bi-directional controller for articulated vehicles," Proceedings of the HVTT15, 2018.

[31] J. Morales, A. Mandow, J. L. Martinez, J. L. Martínez, and A. J. García-Cerezo, "Driver assistance system for backward maneuvers in passive multi-trailer vehicles," in *2012 IEEE/RSJ International Conference on Intelligent Robots and Systems*, Oct. 2012, pp. 4853–4858. doi: 10.1109/IROS.2012.6385799.

[32] E. Kayacan, E. Kayacan, H. Ramon, and W. Saeys, "Distributed nonlinear model predictive control of an autonomous tractor–trailer system," *Mechatronics*, vol. 24, no. 8, pp. 926–933, Dec. 2014, doi: 10.1016/j.mechatronics.2014.03.007.

[33] Y. Sklyarenko, F. Schreiber, and W. Schumacher, "Maneuvering assistant for truck and trailer combinations with arbitrary trailer hitching," in *2013 IEEE International Conference on Mechatronics (ICM)*, Feb. 2013, pp. 774–779. doi: 10.1109/ICMECH.2013.6519139.

[34] A. C. Manav and I. Lazoglu, "A Novel Cascade Path Planning Algorithm for Autonomous Truck-Trailer Parking," *IEEE Trans. Intell. Transp. Syst.*, vol. 23, no. 7, pp. 6821–6835, Jul. 2022, doi: 10.1109/TITS.2021.3062701.

[35] P. Zips, M. Böck, and A. Kugi, "An optimisation-based path planner for truck-trailer systems with driving direction changes," in *2015 IEEE International Conference on Robotics and Automation (ICRA)*, May 2015, pp. 630–636. doi: 10.1109/ICRA.2015.7139245.

[36] A. Mohamed, J. Ren, H. Lang, and M. El-Gindy, "Optimal collision free path planning for an autonomous articulated vehicle with two trailers," in *2017 IEEE International Conference on Industrial Technology (ICIT)*, Mar. 2017, pp. 860–865. doi: 10.1109/ICIT.2017.7915472.

[37] O. Ljungqvist, N. Evestedt, D. Axehill, M. Cirillo, and H. Pettersson, "A path planning and path-following control framework for a general 2-trailer with a car-like tractor," *J. Field Robot.*, vol. 36, no. 8, pp. 1345–1377, 2019, doi: 10.1002/rob.21908.

[38] K. Kvarnfors, "Motion Planning for Parking a Truck and Trailer System," Thesis, School of Electrical Engineering and Computer Science, KTH Royal Institute of Technology, Stockholm, Sweden, 2019.

[39] D. Zobel, E. Balcerak, and T. Weidenfeller, "Minimum Parking Maneuvers for Articulated Vehicles with One-Axle Trailers," in *Robotics and Vision 2006 9th International Conference on Control, Automation*, Dec. 2006, pp. 1–6. doi: 10.1109/ICARCV.2006.345243.

[40] P. Svestka and J. Vleugels, "Exact motion planning for tractor-trailer robots," in *Proceedings of 1995 IEEE International Conference on Robotics and Automation*, May 1995, pp. 2445–2450 vol.3. doi: 10.1109/ROBOT.1995.525626.

[41] X. Zhang, J. Eck, and F. Lotz, "A Path Planning Approach for Tractor-Trailer System based on Semi-Supervised Learning," in *2022 IEEE 25th International Conference on Intelligent Transportation Systems (ITSC)*, Oct. 2022, pp. 3549–3555. doi: 10.1109/ITSC55140.2022.9922552.

[42] G. Lei and Y. Zheng, "Research on Cooperative Trajectory Planning Algorithm Based on Tractor-Trailer Wheeled Robot," *IEEE Access*, vol. 10, pp. 64209–64221, 2022, doi: 10.1109/ACCESS.2021.3062392.

[43] "fmincon - Find minimum of constrained nonlinear multivariable function - MATLAB." Website. Accessed: Jun. 04, 2025. [Online]. Available: https://www.mathworks.com/help/optim/ug/fmincon.html

[44] L. Guvenc, B. Aksun-Guvenc, S. Zhu, and S. Y. Gelbal, *Autonomous Road Vehicle Path Planning and Tracking Control*. John Wiley & Sons, 2021.